\pdfoutput=1
\documentclass[preprint,12pt,authoryear]{elsarticle}

\usepackage{amssymb}
\usepackage{amsmath}
\usepackage{siunitx}
\usepackage{subcaption}
\usepackage{multirow}
\usepackage[table]{xcolor}

\journal{Journal of Theoretical Biology}

\begin{document}

\abovedisplayskip=12pt
\belowdisplayskip=12pt
\setlength{\jot}{2ex}

	\begin{frontmatter}

		\title{A Two-Phase Model of Early Fibrous Cap Formation in Atherosclerosis}
		\author[syd]{Michael G. Watson\corref{cor1}}
		\author[oxf]{Helen M. Byrne}
		\author[syd]{Charlie Macaskill}
		\author[syd]{Mary R. Myerscough}
		\cortext[cor1]{Corresponding author: michael.watson@sydney.edu.au}
		\address[syd]{School of Mathematics and Statistics, University of Sydney, Australia}
		\address[oxf]{Wolfson Centre for Mathematical Biology, Mathematical Institute, University of Oxford, United Kingdom}

		\begin{abstract}
Atherosclerotic plaque growth is characterised by chronic inflammation that promotes accumulation of cellular debris and extracellular fat in the inner artery wall. This material is highly thrombogenic, and plaque rupture can lead to the formation of blood clots that occlude major arteries and cause myocardial infarction or stroke. In advanced plaques, vascular smooth muscle cells (SMCs) migrate from deeper in the artery wall to synthesise a cap of fibrous tissue that stabilises the plaque and sequesters the thrombogenic plaque content from the bloodstream. The fibrous cap provides crucial protection against the clinical consequences of atherosclerosis, but the mechanisms of cap formation are poorly understood. In particular, it is unclear why certain plaques become stable and robust while others become fragile and vulnerable to rupture.

We develop a multiphase model with non-standard boundary conditions to investigate early fibrous cap formation in the atherosclerotic plaque. The model is parameterised using a range of \emph{in vitro} and \emph{in vivo} data, and includes highly nonlinear mechanisms of SMC proliferation and migration in response to an endothelium-derived chemical signal. We demonstrate that the model SMC population naturally evolves towards a steady-state, and predict a rate of cap formation and a final plaque SMC content consistent with experimental observations in mice. Parameter sensitivity simulations show that SMC proliferation makes a limited contribution to cap formation, and highlight that stable cap formation relies on a critical balance between SMC recruitment to the plaque, SMC migration within the plaque and SMC loss by apoptosis. The model represents the first detailed \emph{in silico} study of fibrous cap formation in atherosclerosis, and establishes a multiphase modelling framework that can be readily extended to investigate many other aspects of plaque development.

		\end{abstract}

		\begin{keyword}
Vascular disease \sep Multiphase \sep Smooth muscle cell \sep Platelet-derived growth factor
		\end{keyword}

	\end{frontmatter}


	\section{Introduction}
Atherosclerosis is an inflammatory disease characterised by the formation of fatty lesions in the inner lining of the artery. These lesions develop slowly over many years and can eventually rupture to cause myocardial infarction or stroke \citep{Hans06}. Vascular smooth muscle cells (SMCs) contribute to the prevention of plaque rupture by synthesising a cap of fibrous tissue that stabilises dangerous plaques \citep{Vass13}. A complete understanding of how SMCs develop and then maintain the fibrous cap is therefore crucial, but experimental investigations are hampered by the extremely slow nature of \emph{in vivo} cap formation. Mathematical modelling provides an attractive alternative to laboratory studies, and in this paper we develop a model to investigate the SMC dynamics that lead to formation of the protective fibrous cap.

The arterial wall consists of a series of distinct tissue layers (Figure~1). A thin sheet of endothelial cells (the endothelium) makes up the innermost layer of the wall. These cells are directly exposed to the flow of blood in the vessel lumen. Immediately beneath the endothelium lies a narrow tissue layer called the intima, which is separated from the underlying media by a dense tissue membrane known as the internal elastic lamina (IEL). Beyond the media there is a further layer known as the adventitia. Growth of an atherosclerotic plaque is triggered when blood-borne low density lipoproteins (LDL) penetrate the vessel endothelium, become oxidised or modified in other ways and accumulate in the intima. The subsequent immune response causes monocytes to exit the bloodstream, differentiate into macrophages and consume the LDL to become foam cells. Death of these lipid-laden macrophages within the vessel wall leads to the deposition of cellular debris and extracellular fat. This stimulates further macrophage recruitment and chronic inflammation typically ensues. The associated accumulation of material (some of which may eventually become necrotic) distends the intima, leading to mechanical disruption and tissue remodelling of the artery \citep{Ross99}.

	\begin{figure}[h!]
		\centering		
  		\includegraphics[height=5cm]{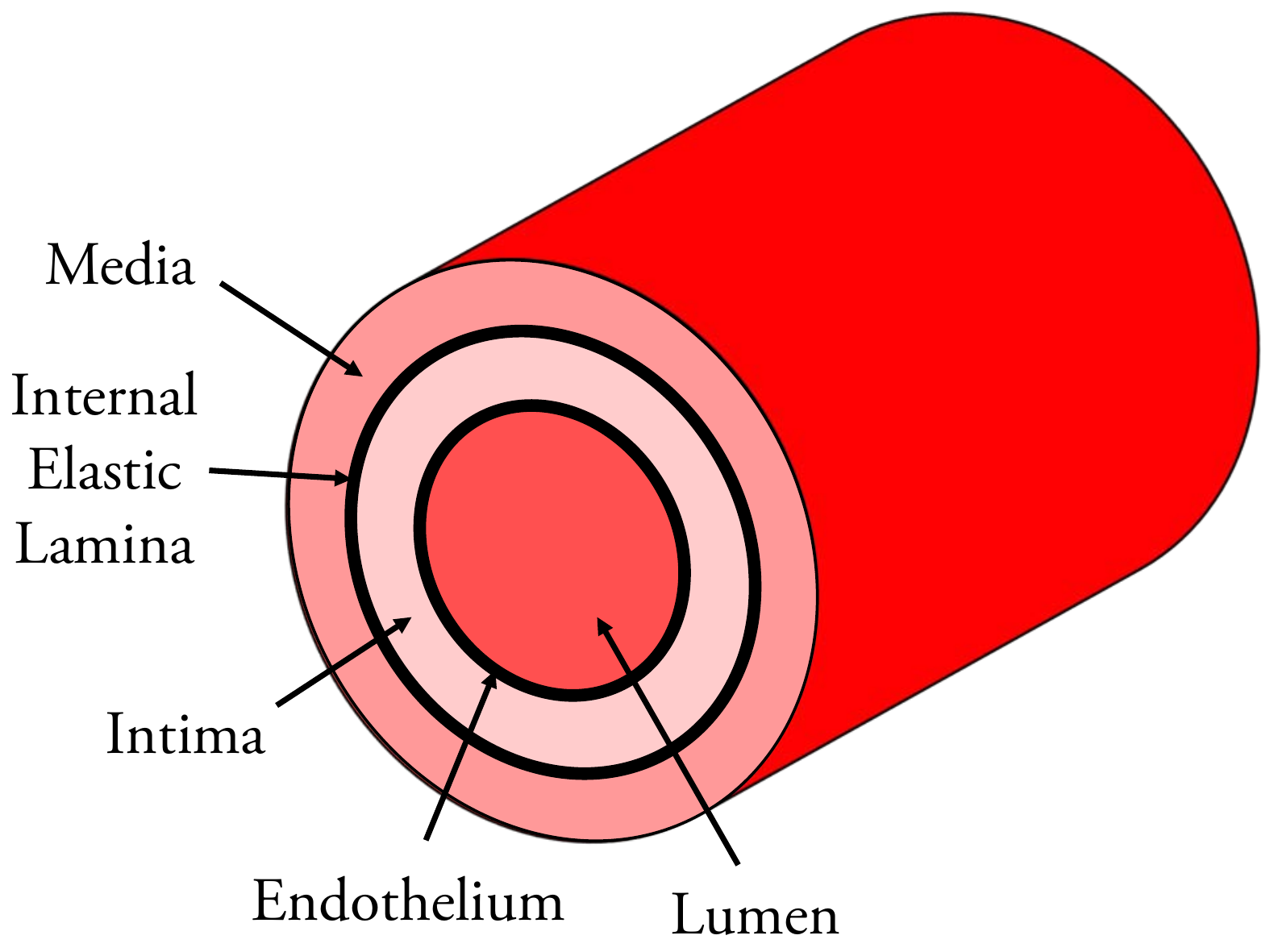}
		\caption{Cross-section of a healthy artery showing the layered structure of the inner artery wall (layer widths not to scale, and adventitia not shown). Atherosclerotic plaques are initiated by infiltration of lipoproteins and immune cells across the endothelium into the intima from the bloodstream in the lumen. In advanced stage plaques, SMCs from the media are activated and migrate through the internal elastic lamina into the intima.}
	\end{figure}

In healthy arteries, most vascular SMCs are in a quiescent (contractile) state within the medial layer of the vessel wall. During atherosclerotic plaque formation, however, these cells are stimulated to adopt a synthetic phenotype and migrate through the IEL to remodel the extracellular matrix (ECM) in the intima \citep{Libb06}. \citet{Fagg84} have suggested that disruption of the endothelium by foam cell accumulation may prompt the activation of SMCs, but the precise mechanisms that underlie the initiation of this response remain unclear. The resulting deposition of a cap of fibrous tissue between the endothelium and the lipid-rich plaque core suggests a key role for SMC chemoattractants in the process of vessel wall remodelling. A variety of cytokines and growth factors have been identified in advanced plaques, but platelet-derived growth factor (PDGF) --- a potent stimulator of both SMC migration and mitosis --- is likely to be the dominant chemoattractant in fibrous cap formation \citep{Ruth97,Sano01}. Indeed, PDGF is known to be expressed by a variery of plaque cells including both endothelial cells and platelets from the bloodstream that adhere to sites of vascular injury \citep{Koza02}. Figure~2 summarises the cap formation process in a schematic diagram.

	\begin{figure}[h!]
		\centering		
  		\includegraphics[height=6cm]{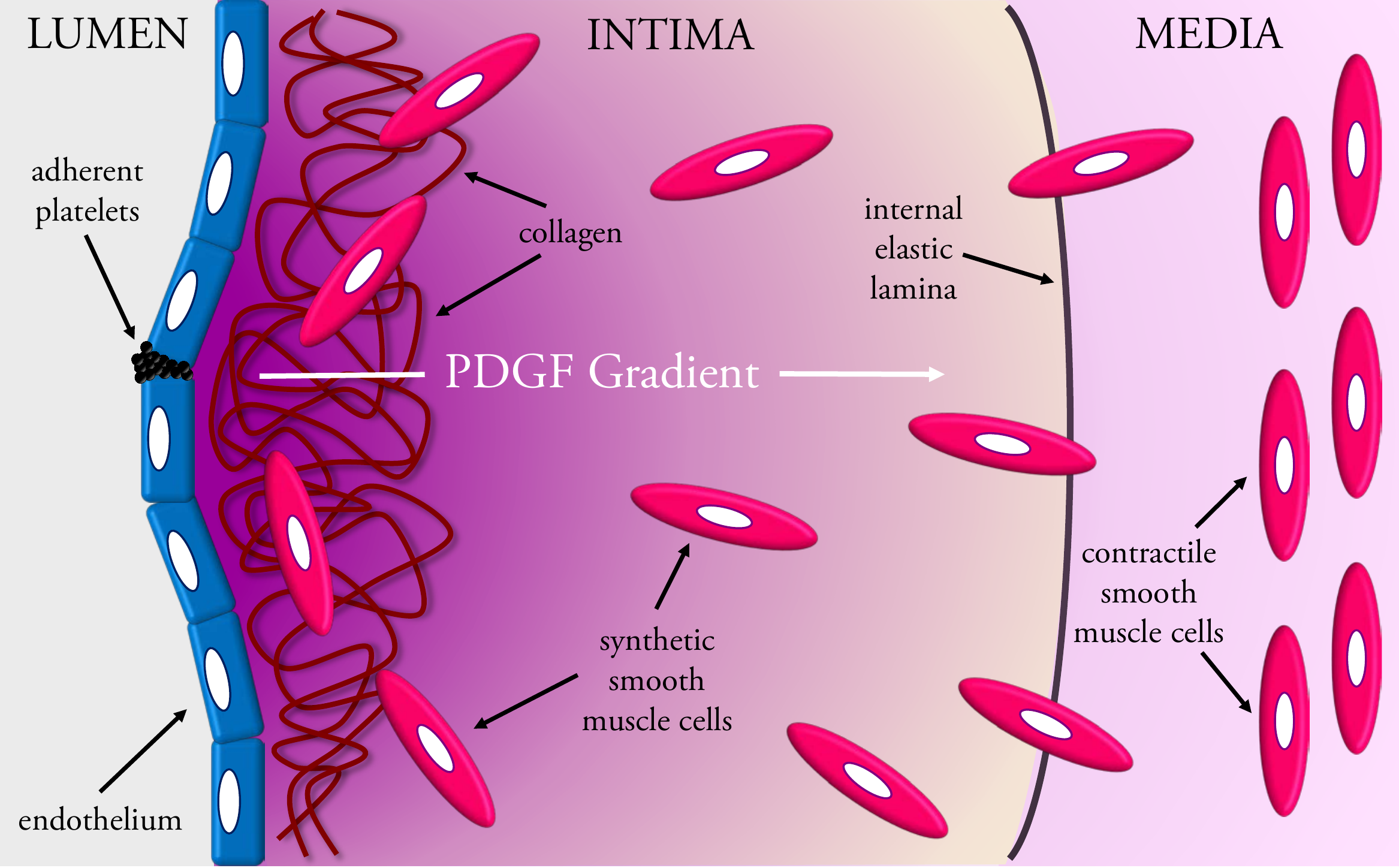}
		\caption{Schematic diagram of the key processes in fibrous cap formation.  The accumulation of lipoproteins and immune cells in early atherosclerotic plaques (not shown) causes expansion of the intima and disruption of the endothelium. Endothelial cells and adherent platelets at sites of endothelial disruption release platelet-derived growth factor (PDGF), which creates a PDGF gradient across the artery wall. Vascular SMCs, which usually reside in a contractile state in the medial layer, respond chemotactically to the PDGF gradient and migrate into the intima through the internal elastic lamina. These so-called synthetic SMCs remodel the ECM in the intima and deposit a cap of collagenous tissue adjacent to the endothelium. This tissue cap provides a dense barrier that prevents the release of thrombogenic plaque material into the bloodstream.}
	\end{figure}

The formation of an atherosclerotic plaque is typically preceded by a gradual expansion of the intima known as diffuse intimal thickening. This phenomenon, which begins in infancy and continues throughout life, involves the deposition of a matrix of elastin and proteoglycans by a small population of non-proliferative SMCs \citep{Naka08}. During plaque development, the existing intimal tissue undergoes substantial changes: migrating media-derived SMCs degrade the existing ECM components and deposit a new matrix composed primarily of fibrillar collagens \citep{Adig09}. Transforming growth factor (TGF)-$\beta$, an anti-inflammatory and pro-fibrotic cytokine that is expressed by a wide variety of plaque cells \citep{Toma12}, is an important mediator of this remodelling process \citep{Lutg02}.

The fibrous collagen cap synthesised by SMCs provides a dense tissue barrier that sequesters thrombogenic plaque material from the bloodstream and contributes to the prevention of plaque rupture. The ability to generate, and then maintain, a cap of sufficient thickness is therefore a key determinant of long-term disease prognosis. The exact mechanisms that underlie fibrous cap development are not well understood, and it remains unclear why certain fibrous caps remain stable and robust while others are fragile and vulnerable to rupture \citep{Virm00, Hans15}. It is frequently posited in the literature that an established fibrous cap can be eroded over time \citep{Lusi00, Libb02}. A number of factors have been linked to this phenomenon --- amplified matrix metalloproteinase (MMP) production due to excessive inflammatory cell recruitment, for example --- but a complete picture has yet to emerge \citep{Newb99}. In this work, we investigate potential sources of plaque instability by developing a mathematical model to study the cellular and biochemical mechanisms that contribute to fibrous cap synthesis. By focussing on the initial SMC response, we address the fundamental question of whether certain atherosclerotic plaques may be pre-disposed to the formation of a thin fibrous cap. This foundational study provides a framework that can ultimately be extended to more detailed investigations of atherosclerotic plaque development.

A variety of mathematical models have been developed to investigate SMC migration and neo-intima formation in the vessel wall. An area of particular research interest is arterial restenosis, which can occur in response to surgical procedures such as angioplasty and stenting. These procedures are used to widen arteries constricted by plaque growth, but they can also damage the endothelium and elicit a response from SMCs that leads to neo-intima formation and recurrence of the luminal narrowing. The SMC behaviour that leads to this restenosis has been modelled using both discrete \citep{Boyl11, Tahi15} and continuum \citep{Lall06} approaches. Modelling has also been performed to study the response of vascular SMCs to other surgical interventions, including vein grafting \citep{Budu08, Garb13} and the implant of blood filters \citep{Nico15}.  Activation of vascular SMCs to repair the vessel wall after endothelial injury has several similarities with dermal wound healing, which has been widely studied in the modelling literature \citep{Mach10, Meno12, Bowd14}. The equivalent cell to the SMC in the wound healing context is the fibroblast, which is recruited to the wound site by growth factors such as PDGF to synthesise new ECM and stimulate wound contraction.  The chemotactic response of fibroblasts to PDGF during wound healing has been modelled in detail by \citet{Haug06}, while a host of approaches have been used to study the re-generation of dermal fibrous tissue by this motile cell population \citep{Olse95, Dale97, Dall98, McDo06, Murp11}.

Mathematical modelling of the cellular activity in the vessel wall during atherosclerosis is an area of growing research interest (see \citet{Part16} for a comprehensive review). To date, however, the vast majority of published models have investigated the actions of monocytes/macrophages in disease initiation \citep{ElKh07, Papp08, Cohe14, Chal15, Bhui17} and only a handful have explicitly considered the role of SMCs in plaque progression. This possibility was first discussed in the work of \citet{McKa04}, who proposed a simple reaction-diffusion framework for plaque formation that included both SMC migration and collagen remodelling. \citet{Cill14} included differentiation of SMCs from a contractile to a synthetic phenotype in their coupled model of blood flow and plaque growth. As with a number of other studies \citep{Ibra05, Post07}, however, the simplified representation of SMC behaviour in this model fails to adequately describe the process of cap formation. \citet{Fok12} used a moving boundary model to investigate mechanisms of intimal thickening. This model included SMC migration and proliferation in response to an endothelium-derived source of PDGF, and is, to the best of our knowledge, the only existing approach that shares similarities with the current work.

Previous work in our group \citep{Ougr10} has seen the development of a detailed reaction-diffusion model describing the spatio-temporal evolution of the key cytokines and tissue constituents involved in long-term plaque development. This model successfully predicted morphological aspects of plaques and led to many interesting outcomes, but certain features of the approach remain unsatisfactory. A particular shortcoming of the model is that it fails to account for volume filling or exclusion as material accumulates in the intima over time. This is an important consideration, and we address this in the current study by modelling fibrous cap formation in the intima using a multiphase approach. Multiphase models, which provide a natural framework to study heterogeneous tissues, have been widely applied in contexts such as cancer \citep{Prez09, Hubb13} and tissue engineering \citep{Lemo06, Pear14}. The approach has also recently been applied in atherosclerosis to study the influence of the plasma lipid profile on plaque growth and regression \citep{Hao14, Frie15}. SMC migration and ECM remodelling were included in this study, but formation of the fibrous cap was not explicitly investigated.

The model of \citet{Hao14} uses fifteen different variables to represent various chemical, cellular and tissue components implicated in atherosclerotic plaque formation. In the current study, we adopt an alternative approach and focus on the interaction between only two key variables. The benefit of this approach is that we can perform a detailed investigation of the cell behaviour and obtain insight that could not be easily achieved using larger models with generalist, non-specific assumptions. Our model considers only the early SMC response to injury-induced chemical signalling from the endothelium and neglects the immune response that drives initial lesion growth. We do not explicitly consider SMC remodelling of the ECM or, indeed, fibrous cap synthesis --- instead we investigate the mechanisms of SMC migration and proliferation in the plaque and assess the likely implications for cap stability.

Full details of the modelling assumptions and the model parameterisation are introduced in the next section. This is followed by simulation results and a parameter sensitivity study. We conclude with a discussion of the outcomes of this work in the context of both atherosclerosis modelling and biomedical atherosclerosis research.

\section{Model Formulation and Parameterisation}

\subsection{Model Description}

We model this system with a multiphase approach on a one-dimensional Cartesian domain $x\in[0, L]$.  This region, which ranges from the endothelium ($x = 0$) to the SMC-bearing media ($x = L$), is assumed to represent a cross-section of diseased intimal tissue far from the edges of the plaque. The composition of the intima evolves in a complex manner during atherosclerosis progression, and the relative quantities of constituents such as lipids, monocytes, foam cells, SMCs, interstitial fluid and ECM proteins change over time. Although each of these elements may ultimately have some role to play in cap formation, the complexity of a multiphase model that comprises all components individually is less likely to provide mechanistic insights than a simpler model. We therefore propose a model that involves only two phases: (1) SMCs and (2) all other cells and tissues (i.e. a non-SMC phase), with volume fractions denoted by $m(x,t)$ and $w(x,t)$, respectively. We do not explicitly consider any remodelling of the ECM within the vessel wall; rather, we assume that, over time, the distribution of deposited matrix --- and thus any fibrous cap --- will evolve to broadly reflect the underlying SMC profile. The variable $m(x,t)$ should, therefore, be defined more specifically as the volume fraction of \emph{matrix-synthesising SMCs}. This is a key assumption. Vascular SMCs are known to be extremely plastic and capable of demonstrating a broad range of behaviours \citep{Owen04, Maje16}, but we will consider only this particular phenotype.

We define a number of additional variables before we generate the model equations. The velocities of the SMC and non-SMC phases are $v_m(x,t)$ and $v_w(x,t)$, and the corresponding stresses experienced by these phases are $\tau_m(x,t)$ and $\tau_w(x,t)$. We also introduce $p(x,t)$ to represent the interstitial fluid pressure. The process of medial SMC activation and subsequent fibrous cap formation relies heavily on a range of chemical cues, with PDGF believed to perform a crucial role as a potent SMC chemoattractant \citep{Ross90, Newb99}. Therefore, in addition to the volume-occupying phases within the intima, we include a profile of diffusible PDGF. This is assumed to be present only in the non-SMC phase, with concentration $c(x,t)$.

\subsubsection*{Mass Balance Equations}

Given that biological tissues are primarily composed of water, we assume that both the SMC and non-SMC phases have the same constant density.  Applying mass balances to each phase, we obtain:
	\begin{align}
		\frac{\partial m}{\partial t} + \frac{\partial}{\partial x}\left(v_m m\right) &= S_m\left(c,m,w\right), \\
		\frac{\partial w}{\partial t} + \frac{\partial}{\partial x}\left(v_w w\right) &= -S_m\left(c,m,w\right),
	\end{align}
where $S_m$ denotes the net rate of matrix-producing SMC proliferation or loss within the intima, which may depend on $c$, $m$ and $w$. Note that we have also assumed that there is no local source or sink of material: proliferating cells are assumed to grow by absorbing material (e.g. water) from the surrounding tissue, while cells that die (or simply cease ECM synthesis) return to the bulk tissue phase. As suggested by the definitions of $m$ and $w$, there is a no-voids condition:
	\begin{equation}
		m + w = 1,
	\end{equation}
which reflects an underlying assumption that the intimal tissue within the vessel wall is both continuous and incompressible.

The characteristic timescale of chemical diffusion of PDGF is likely to be significantly shorter than that of SMC migration. Therefore, we neglect the role of advective transport within the non-SMC phase, and assume that the concentration of PDGF in the intima is at steady-state and given by:
	\begin{equation}
		D_c \, \frac{\partial}{\partial x}\left(w \, \frac{\partial c}{\partial x}\right) = \eta_c m w c + \beta_c w c,
	\end{equation}
where $\eta_c$ and $\beta_c$ represent the SMC uptake and chemical decay rates, respectively \citep{Asta08}. Note that the reaction terms both include a dependence on $w$ under the assumption that neither uptake nor decay occurs in the absence of non-SMC material. This is because the model assumes that PDGF is only present in the non-SMC fraction. The diffusion coefficient of PDGF is $D_c$ and $w$ modulates diffusive transport under the assumption that limited transport occurs when the SMC volume fraction is high or, equivalently, $w$ is low.

There is no source term for PDGF in the above equation. PDGF is introduced into the domain through an appropriate boundary condition at the lumen. We postpone the detailed discussion of boundary conditions until later, but we emphasise here the crucial role that boundary conditions will play in capturing many of the key aspects of cap formation.

\subsubsection*{Momentum Balance Equations and Constitutive Assumptions}

In the absence of external body forces, and assuming that inertial effects are negligible, the momentum equations reduce to a balance between intraphase and interphase forces for the SMC and non-SMC phases, respectively:
	\begin{align}
		\frac{\partial}{\partial x}\left(\tau_m m\right) &= -F_{mw}, \\
		\frac{\partial}{\partial x}\left(\tau_w w\right) &= F_{mw}.
	\end{align}
The interphase force exerted by the non-SMC phase on the SMC phase is denoted by $F_{mw}$ (note that the SMCs exert an equal and opposite force on the other phase), and we assume an expression of the following form \citep{Brew02, Byrn04}:
	\begin{equation}
		F_{mw} = -p \, \frac{\partial w}{\partial x} + k m w \left(v_w - v_m\right).
	\end{equation}
The first term on the right hand side arises from the volumetric averaging process that underpins the multiphase approach \citep{Drew71}. It contributes a force due to the distribution of pressure at the interface between the phases. On the microscale, the force experienced by an element of one phase due to the other phase will increase with increasing interfacial contact between the phases.  The form of this term reflects the overall macroscopic manifestation of this microscale phenomenon \citep{Lemo06}.  The second term in equation (7) represents a simple viscous drag between the phases that is characterised by the drag coefficient $k$.

For the respective stress terms $\tau_m$ and $\tau_w$ in equations (5) and (6) above, we propose the following models \citep{Brew02, Byrn04}:
	\begin{align}
		\tau_m &= -p - \Lambda \left(c\right), \\
		\tau_w &= -p,
	\end{align}
where, for simplicity, we have neglected viscous effects within either phase. The non-SMC phase comprises a mixture of cells and tissues that, in reality, is likely to have some influence on SMC migration within the intima. We will assume here that it is an inert isotropic fluid in order to focus primarily on the role played by PDGF in fibrous cap formation. The second term on the right hand side of equation (8) reflects an assumption that SMC behaviour is more complex than that of the non-SMC phase, with cell motility assumed to be influenced by the local PDGF concentration. We define the function $\Lambda \left(c\right)$ to be:
	\begin{equation}
		\Lambda \left(c\right) = \frac{\chi}{1 + {\left(\kappa c \right)}^n},
	\end{equation}
where $\chi$, $\kappa$ and $n$ are positive parameters \citep{Byrn04}. This model assumes that SMC phase pressure decreases with increasing PDGF concentration. This assumption allows movement up the chemical gradient (i.e. chemotaxis) to be represented as a mechanism of stress relief in the SMC phase.

\subsubsection*{Model Simplification}

In this section, we use the above definitions to derive a reduced nonlinear system of PDEs that describes early fibrous cap formation in the intima. The model comprises a single parabolic equation for the SMC volume fraction $m(x,t)$ coupled to an elliptic reaction-diffusion equation for the PDGF concentration $c(x,t)$. We consider the boundary and initial conditions required to close the system in the next section. The model simplification is similar to that reported in \citet{Byrn04}, but we include the full details for clarity.

We use the no-voids condition of equation (3) to replace all occurrences of the non-SMC phase volume fraction $w(x,t)$ by its equivalent expression $1 - m(x,t)$. Under this assumption, the sum of the mass balance equations (1) and (2) reduces to:
	\begin{equation}
		\frac{\partial}{\partial x} \left(v_m m + v_w \left(1 - m \right)\right) = 0.
	\end{equation}
Integrating this equation, and applying a zero-flux condition for SMCs at the lumen (i.e. $v_m = v_w = 0$ at $x = 0$), the velocity $v_w$ of the non-SMC phase can be expressed in terms of the velocity $v_m$ of the SMC phase in the following way:
	\begin{equation}
		v_w = -v_m \left( \frac{m}{1 - m} \right).
	\end{equation}
Using the definitions (8) and (9), the momentum balance equations (5) and (6) can be added to give the relationship:
	\begin{equation}
		\frac{\partial}{\partial x} \left(\tau_m m + \tau_w \left(1 - m \right)\right) = - \frac{\partial}{\partial x} \left(p + m \Lambda \right) = 0.
	\end{equation}
Note that there is no requirement to perform the integration here. We will use this expression in its current form to eliminate the tissue pressure $p(x,t)$ from the equations. For now, we proceed by substituting the expressions (7) and (9) into equation (6) for the non-SMC phase momentum balance to find:
	\begin{equation}
		- \frac{\partial}{\partial x} \left(p \left(1 - m \right)\right) = -p \, \frac{\partial}{\partial x} \left(1 - m \right) + k m \left(1 - m \right) \left(v_w - v_m \right).
	\end{equation}
By expanding the left hand side and cancelling terms, equation (14) can be reduced to:
	\begin{equation}
		- \frac{\partial p}{\partial x} = k m \left(v_w - v_m \right).
	\end{equation}
Now, using equations (13) and (12), respectively, we can eliminate all other variables in equation (15) and express $v_m$ simply in terms of $m$ and $c$:
	\begin{equation}
		v_m = v_w - \frac{1}{k m} \frac{\partial}{\partial x} \left(m \Lambda \right) = -\frac{\left(1 - m \right)}{k m} \frac{\partial}{\partial x} \left(m \Lambda \right).
	\end{equation}
This expression for the cell phase velocity can then be substituted back into equation (1), to derive the final mass balance relationship for SMCs in the intima:
	\begin{equation}
		\frac{\partial m}{\partial t} = \frac{1}{k} \frac{\partial}{\partial x} \left[\left(1 - m \right) \frac{\partial}{\partial x} \left(m \Lambda \left(c \right) \right) \right] + S_m\left(c,m\right).
	\end{equation}
For consistency, we may also remove all references to $w(x,t)$ in equation (4), and our simplified model then comprises equation (17) along with the following expression for the PDGF concentration:
	\begin{equation}
		D_c \, \frac{\partial}{\partial x}\left(\left(1 - m \right) \frac{\partial c}{\partial x}\right) = \eta_c c m \left(1 - m \right) + \beta_c c \left(1 - m \right).
	\end{equation}

\subsubsection*{Boundary and Initial Conditions}

Before discussing the boundary conditions for our system, it is instructive to expand the flux term in equation (17) and separate it into explicit diffusive and chemotactic contributions. We use this format to define an appropriate boundary condition for SMC entry into the intima from the media. Recalling the definition of the extra pressure in equation (10), and applying the chain rule to the flux term, we can write:
	\begin{equation}
		\frac{\partial m}{\partial t} = \frac{\partial}{\partial x} \left[U\left(c,m \right) \frac{\partial m}{\partial x} - V\left(c,m \right) m \, \frac{\partial c}{\partial x} \right] + S_m\left(c,m\right),
	\end{equation}
where the effective, nonlinear diffusion coefficient is given by:
	\begin{equation}
		U\left(c,m \right) = \frac{1}{k} \left(1 - m \right) \Lambda = \frac{1}{k} \left(1 - m \right) \left[\frac{\chi}{1 + {\left(\kappa c \right)}^n}\right],
	\end{equation}
and the effective, nonlinear chemotaxis coefficient is:
	\begin{equation}
		V\left(c,m \right) = -\frac{1}{k} \left(1 - m \right) \frac{d \Lambda}{d c} = \frac{1}{k} \left(1 - m \right) \left[\frac{\chi n \kappa^n c^{n-1}} {\left(1 + \left(\kappa c \right)^n\right)^2}\right].
	\end{equation}
We emphasise here the dependence of the chemotaxis coefficient on $c$, which implies that the overall chemotactic flux at any point in the domain will depend not only on the local PDGF gradient, but also on the particular local PDGF concentration.

Returning to the boundary conditions, we recall an earlier assumption in the derivation of equation (12) that no SMCs will enter the lumen from the intima. Therefore, we have the following condition:
	\begin{align}
		\frac{1}{k} \left(1 - m \right) \frac{\partial}{\partial x} \left(m \Lambda \right) &= 0 \text{ at } x = 0 \nonumber \\
		\implies \frac{\partial} {\partial x} \left(m \Lambda \right) &= 0 \text{ at } x = 0.
	\end{align}
At the medial boundary, we assume that a prior signalling cascade --- most likely involving MMPs --- has allowed the resident (contractile) SMCs to degrade their local tissue and adopt a migratory (synthetic) phenotype \citep{Newb99}. To invade the intima from the media, the SMCs must first pass through the porous IEL, which is a dense tissue layer that separates the intima from the media and provides a physical barrier to cell movement. We therefore assume that migration of SMCs across this boundary is driven entirely by chemotaxis to PDGF, and that the contribution from passive diffusion is negligible. We introduce the parameter $m^*<1$ to represent a constant synthetic SMC volume fraction in the media, and assume that these SMCs enter the intima via a chemotactic response to the PDGF at the boundary, so that:
	\begin{align}
		\frac{1}{k} \left(1 - m \right) \frac{\partial}{\partial x} \left(m \Lambda \right) &= \frac{1}{k} \left(1 - m \right) \frac{d\Lambda}{d c} \, m^* \, \frac{\partial c}{\partial x} \text{ at } x = L \nonumber \\
		\implies \frac{\partial}{\partial x} \left(m\Lambda \right) &= \frac{d\Lambda}{d c} \, m^* \, \frac{\partial c}{\partial x} \text{ at } x = L.
	\end{align}
Note that the format of this boundary chemotactic flux is consistent with the terms derived in equations (19) and (21), and also that this condition does not stipulate continuity of $m$ across the boundary. A further consequence of this boundary condition is that the influx of SMCs into our closed system must be balanced by an equivalent \emph{efflux} from the non-SMC phase. Physically, we may interpret this efflux as a transfer of interstitial fluid towards the medial tissue which has been locally degraded and vacated by activated SMCs. This physical interpretation is inconsistent with the content of the non-SMC phase. We would not anticipate, for example, a net efflux of macrophages out of the intima. However, we believe the model remains reasonable provided that the overall intimal SMC population does not become too large. Indeed, experimental measurements from mouse models indicate typical values of around just 10\% for the total SMC content in plaques \citep{Mall01, Clar06}. Alternative modelling approaches that specify additional phases and/or employ moving boundaries could capture many features of plaque growth more naturally, but our aim here is to develop a simple, foundational model of fibrous cap formation.

As mentioned earlier, we assume that the sole source of SMC-stimulating PDGF is from an inward flux at the lumen.  A number of cell types, including macrophages and SMCs themselves, have been implicated in the production of PDGF during plaque development \citep{Libb06}, but stimulated endothelial cells and adherent platelets are believed to be the most significant sources of the chemical \citep{Funa98}.  We model this with the following boundary condition:
	\begin{align}
		D_c \left(1 - m \right) \frac{\partial c}{\partial x} &= -\alpha_c \left(1 - m \right) \text{ at } x = 0 \nonumber \\
		\implies \frac{\partial c}	{\partial x} &= -\frac{\alpha_c}{D_c} \text{ at } x = 0,
	\end{align}
where, in the absence of SMCs at the lumen, $\alpha_c$ represents a fixed rate of PDGF flux into the intima. The term $\left(1-m\right)$ is included to provide consistency with the form of the nonlinear flux inside the domain; we note, however, the inherent implication that the rate of PDGF supply to the intima will diminish with increasing SMC fraction at the boundary. This is consistent with an assumption that SMCs inhibit, either in a direct or indirect manner, the stimulation of endothelial cell PDGF production at the lumen.

Finally, we define a condition for the PDGF at the medial boundary. Inspection of equation (23) indicates that in order for SMCs to migrate into the intima from the media, a (negative) chemical gradient \emph{must} exist at this point.  To ensure the existence of such a gradient, we treat the IEL as a porous membrane with permeability to PDGF denoted by $\sigma_c$, and assume a diffusion-driven flux of chemical across the boundary:
	\begin{align}
		D_c \left(1 - m \right) \frac{\partial c}{\partial x} &= \sigma_c \left(c^* - c \right) \left(1 - m \right) \text{ at } x = L \nonumber \\ 
		\implies \frac{\partial c}{\partial x} &= \frac{\sigma_c}{D_c} \left(c^* - c \right) \text{ at } x = L.
	\end{align}
Here, $c^*$ represents a notional background PDGF concentration in the media, and we again include a $\left(1-m\right)$ term for consistency with the modulated flux inside the domain.

Since the PDGF concentration is assumed to evolve in a quasi-steady manner, an initial condition is not required to solve equation (18) and we need only to define an initial SMC profile within the intima. In theory, this profile should reflect the SMC population within \emph{healthy} intimal tissue, which is known to vary across species. Small populations of SMCs are believed to reside in the intimal tissue of humans, pigs and primates; however, in the majority of other animal models --- most importantly in the genetically modified mice used as an animal model to study plaque development --- this population is known to be completely absent \citep{Newb99}. In this study we assume that the initial volume fraction of SMCs in the intima is negligible. Since the model equations cannot support a domain entirely devoid of SMCs (see the singularity arising in equation (16), for example), we implement this assumption by assigning a small, uniform initial SMC population:
	\begin{equation}
		m\left(x,0\right) = m_i, \text{ where } 0<m_i \ll 1.
	\end{equation}

\subsubsection*{SMC Phase Source Terms}

The full model comprises equations (17) and (18), subject to the boundary and initial conditions stated in equations (22)-(26). The format of the generic source term $S$ in equation (17) remains to be determined.

Experimental studies have shown that SMCs proliferate throughout the process of plaque formation \citep{Lutg99}. Rates of mitosis have been reported to be exceedingly low \citep{Gord90} but, over the long timescale associated with disease progression, proliferation may still provide a significant source of plaque SMCs \citep{Newb99}. \emph{In vitro} evidence has also indicated the upregulation of SMC mitosis in the presence of increasing PDGF concentration \citep{Munr94}. Therefore, as well as a linear rate of cell loss, our SMC source term includes logistic proliferation with a growth factor-enhanced linear mitosis rate \citep{Olse95}:
	\begin{equation}
		S_m \left(c,m\right) = m \left(1 - m \right) \left(r_{m0} + r_m \left[\frac{c}{c_m + c}\right]\right) - \beta_m m.
	\end{equation}
Here, $\beta_m$ is the rate of cell loss, $r_{m0}$ is the baseline mitosis rate, $r_m$ quantifies the maximum possible PDGF-stimulated increase in the mitosis rate, and $c_m$ represents the PDGF concentration where the half-maximal increase in mitosis occurs.

\subsubsection*{Model Non-Dimensionalisation}

Using tildes to denote dimensionless quantities, we rescale space $x$, time $t$ and PDGF concentration $c$ as follows (note that the SMC volume fraction $m$ does not require rescaling):
	\begin{equation*}
		\widetilde{x} = \frac{x}{L}, \; \widetilde{t} = \frac{t}{t_0}, \; \widetilde{c} = \frac{c}{c_0}, \; \widetilde{m} = m,
	\end{equation*}
where $t_0$ and $c_0$ represent characteristic time and PDGF concentration values, respectively. The model parameters can now be non-dimensionalised in the following way:
	\begin{equation*}
		\begin{gathered}
			\widetilde{\chi} = \frac{\chi t_0}{k L^2}, \; \widetilde{\kappa} = \kappa c_0, \; \widetilde{r} = r_{m0} t_0, \; \widetilde{A} = \frac{r_m}{r_{m0}}, \; \widetilde{c_m} = \frac{c_m}{c_0}, \; \widetilde{\beta_m} = \beta_m t_0, 						\\
			\widetilde{\eta_c} = \frac{\eta_c L^2}{D_c}, \; \widetilde{\beta_c} = \frac{\beta_c L^2}{D_c}, \; \widetilde{\alpha_c} = \frac{\alpha_c L}{D_c c_0}, \; \widetilde{\sigma_c} = \frac{\sigma_c L}{D_c}, \; 		\widetilde{c^*} = \frac{c^*}{c_0}.
		\end{gathered}
	\end{equation*}
Dropping tildes for clarity, the corresponding dimensionless model equations, boundary conditions and initial conditions are:
	\begin{gather}
		\frac{\partial m}{\partial t} = \frac{\partial}{\partial x} \left[\left(1 - m \right) \frac{\partial}{\partial x} \left(m \Lambda \right) \right] + r m \left(1 - m \right) \left(1 + \frac{A c}{c_m + c}\right) - \beta_m m, \\
		\frac{\partial}{\partial x}\left[\left(1 - m \right) \frac{\partial c}{\partial x}\right] = \eta_c c m \left(1 - m \right) + \beta_c c \left(1 - m \right), \\
		\frac{\partial}{\partial x} \left(m \Lambda \right) = 0 \text{ at } x = 0, \; \frac{\partial}{\partial x} \left(m\Lambda \right) = \frac{d\Lambda}{d c} \, m^* \, \frac{\partial c}{\partial x} \text{ at } x = 1, \\
		\frac{\partial c}{\partial x} = -\alpha_c \text{ at } x = 0, \; \frac{\partial c}{\partial x} = \sigma_c \left(c^* - c \right) \text{ at } x = 1, \\
		m\left(x,0\right) = m_i,
	\end{gather}
wherein $\Lambda \left(c\right) = \dfrac{\chi}{1 + {\left(\kappa c \right)}^n}$.

\subsection{Model Parameterisation}

The majority of our parameters have been informed by data from previous computational and experimental studies. When appropriate data could not be obtained, we have selected values that ensure biologically reasonable results. See Table 1 for the base case parameter set.

	\begin{table}[h!]
		\centering
		\footnotesize
		\begin{tabular}{| c | p{4.8cm} | c | p{3.2cm} |} 
			\hline
			\multirow{2}{*}{Parameter} & \multirow{2}{*}{Description} & Dimensionless & \multirow{2}{*}{Reference} \\
			& & Value & \\ \hline
			$n$ & Exponent in SMC phase pressure function & 1.8 & \citet{Scha97} \\ \hline
			$\kappa$ & Reciprocal of reference PDGF concentration in SMC phase pressure function & 5.5 & \citet{Scha97} \\ \hline
			$\chi$ & SMC motility coefficient & 4 & \citet{Cai07} \\ \hline
			$\beta_m$ & Rate of synthetic SMC loss & 0.4 & \citet{Lutg99} \\ \hline
			$\eta$ & Rate of PDGF uptake by SMCs & 1 & \\ \hline
			$\beta_c$ & Rate of PDGF decay & 0.075 & \citet{Haug06} \\ \hline
			$\alpha_c$ & Rate of PDGF influx from endothelium & 0.55 & \\ \hline
			$\sigma_c$ & Permeability of IEL to PDGF & 2 & \\ \hline
			$m^*$ & Volume fraction of synthetic SMCs in media & 0.01 & \\ \hline
			$r$ & Baseline rate of SMC proliferation & 0.02 & \citet{Bret86} \citet{Lutg99} \\ \hline
			$c^*$ & PDGF concentration in media & 0 & \\ \hline
			$A$ & Maximal factor of PDGF-stimulated increase in rate of SMC proliferation & 19 & \citet{Munr94} \\ \hline
			$c_m$ & PDGF concentration for half-maximal increase in SMC proliferation & 1.4 & \citet{Munr94} \\ \hline
			$m_i$ & Initial SMC volume fraction in intima & $1.0\times{10}^{-4}$ & \\
			\hline
		\end{tabular}
		\caption{Base case parameter values. The final column highlights any references that have been used to calculate individual parameter values. The values of parameters without references have been chosen to ensure biologically reasonable results. Unless otherwise stated, all reported simulations use these values.}
	\end{table}

The progression of atherosclerosis is known to be slow, so we assume a characteristic timescale of approximately $1$ month by setting $t_0=$~\SI{2.5e6}{\second}. The domain length $L$, which represents a typical intimal width at the time SMC migration from the media is triggered, is chosen to be \SI{75}{\micro\metre} \citep{Groo15}. Using cell trajectory data from \emph{in vitro} experiments, \citet{Cai07} estimated the diffusion coefficient of fibroblasts to range from \SI{5.0e-3}{\micro\metre\squared\per\second} to \SI{5.0e-2}{\micro\metre\squared\per\second} depending on the extent of cell crow- ding. To account for the influence of the macrophages and large foam cells that also occupy the plaque tissue, we choose the SMC diffusion coefficient near the middle of this range. The spatial and temporal scaling values give the dimensionless motility coefficient $\chi = 4$.

For the PDGF, we assume a characteristic physiological concentration $c_0 =$~\SI{10}{\nano\gram\per\milli\litre}. This is consistent with estimates made by \citet{Olse95} based on reported concentrations of PDGF storage in platelets (\numrange[range-phrase = --]{15}{50}~\si{\nano\gram\per\milli\litre}; \citet{Huan88}). Diffusion coefficient and decay rate values from \citet{Haug06} suggest a PDGF diffusion distance of around \SI{300}{\micro\metre}. In line with this, we set the dimensionless PDGF decay rate $\beta_c=0.075$. Estimates for $\alpha_c$ and $\sigma_c$ are more difficult to obtain, so we select values that give a reasonable initial profile of PDGF within the intima. Note that we also set $c^*=0$, since the media is not known to provide a significant source of PDGF.

The invasion of the intima by SMCs in our model is driven entirely by the presence of a PDGF gradient. This migration ultimately depends on the function $\Lambda$ and its negated derivative $-\frac{d \Lambda}{d c}$, which are proportional to the SMC diffusion and chemotaxis coefficients, respectively. Obtaining reasonable estimates for the parameters $\kappa$ and $n$ is therefore crucial, and we do this by using \emph{in vitro} data on the chemotactic response of human vascular SMCs to PDGF (\citet{Scha97}; see Figure~3). The experimental results predict maximal chemotaxis at a PDGF concentration around \SI{1}{\nano\gram\per\milli\litre}, with a gradual drop-off for lower PDGF levels and a sharper drop-off for higher PDGF levels. By setting $\kappa=5.5$ and $n=1.8$, we obtain a prediction for the form of $\Lambda$ and an excellent fit for $-\frac{d \Lambda}{d c}$.

	\begin{figure}[h!]
		\centering		
  		\includegraphics[height=5cm]{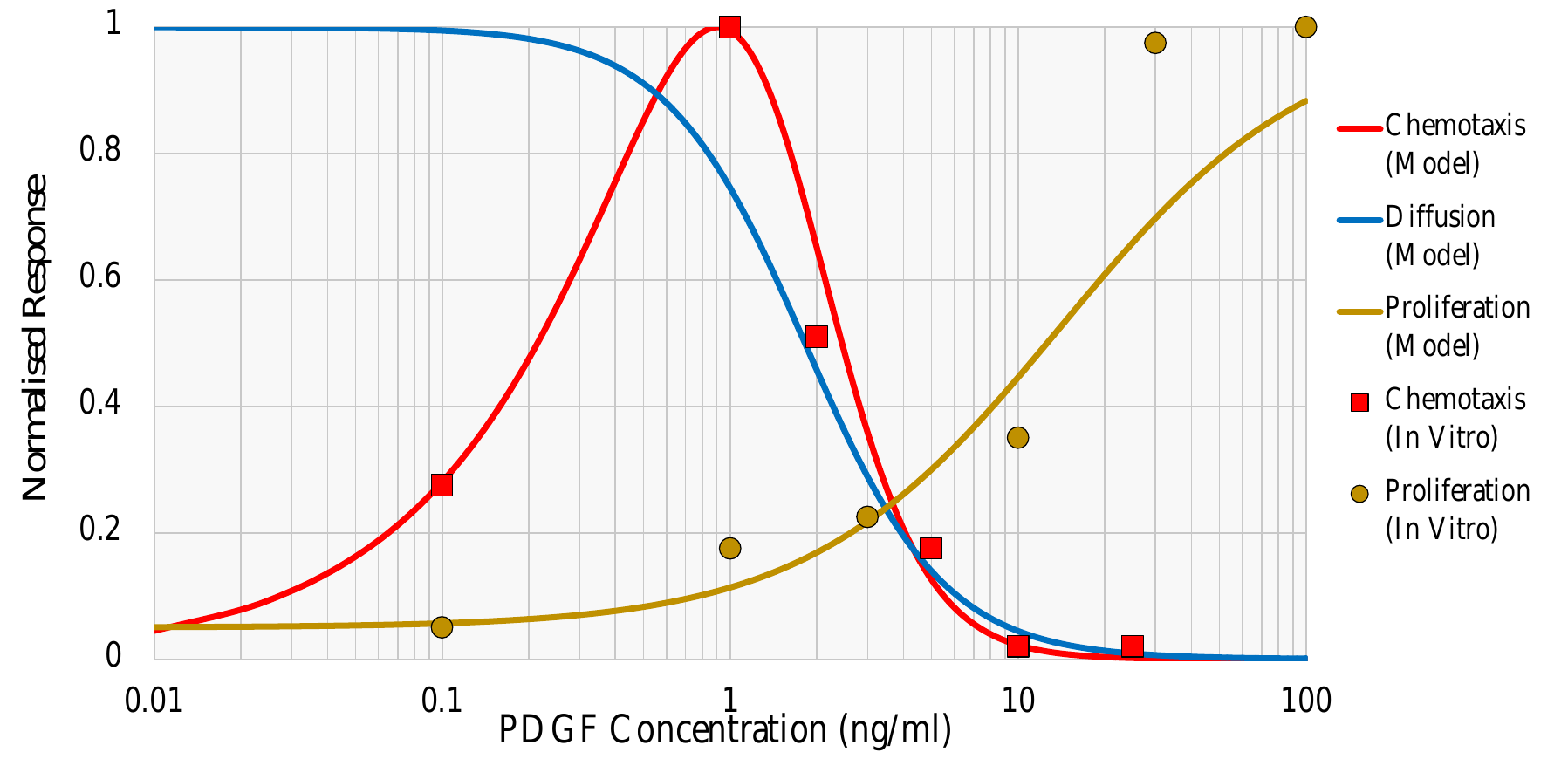}
		\caption{Responses of SMCs across a range of PDGF concentrations in the model (solid lines) and from \emph{in vitro} experiments (data points).  The blue line, which corresponds to the SMC phase pressure function $\Lambda\left(c\right)$ (see equation (10)), shows the variation in the normalised SMC diffusion coefficient with PDGF concentration.  The red line, which corresponds to $-\frac{d \Lambda}{d c}$, shows the variation in the normalised SMC chemotaxis coefficient with PDGF concentration.  The function $\Lambda\left(c\right)$ has been parameterised to fit the experimental data on SMC chemotaxis (red squares) using parameter values $n = 1.8$ and $\kappa = 5.5$.  The gold line shows the normalised variation in SMC proliferation with PDGF concentration, and has been fitted to the experimental data (gold circles) using a simple functional form (see equation (28)) with parameter values $A = 19$ and $c_m = 1.4$.  Proliferation data was obtained from \citet{Munr94} and chemotaxis data was obtained from \citet{Scha97}.}
	\end{figure}

Using further experimental data, we also obtain estimates for the parameters in the SMC source terms. \emph{In vitro} measurements from \citet{Munr94} on the proliferative response of SMCs to PDGF are plotted in Figure~3. Fitting these to our PDGF-enhanced mitosis function suggests the parameter values $A = 19$ and $c_m = 1.4$. For PDGF concentrations in the physiologically realistic range (i.e.\ around \SI{10}{\nano\gram\per\milli\litre} and below), this simple functional form provides a good fit to the experimental data.

The parameters $A$ and $c_m$ define the manner in which PDGF stimulates SMC division in the model, but not the actual rate of proliferation. To estimate this rate in the intima during fibrous cap development we look instead to \emph{in vivo} measurements. Proliferative indices for cells (i.e.\ the observed fraction of cells undergoing mitosis at a fixed time point) in advanced human atherosclerotic plaques have been calculated to be no more than 1\% \citep{Gord90, Lutg99}. Taking a typical SMC doubling time to be 32 hours \citep{Bret86}, and adjusting downwards for potential PDGF-stimulated proliferation in these measured indices, we assume a baseline proliferation rate $r = 0.02$. \citet{Lutg99} also identified apoptotic indices in advanced plaques to be around 1\%. Since cell apoptosis is likely to be much quicker than cell division, we assume the rate of SMC death to be at least an order of magnitude higher than that for proliferation and set $\beta_m = 0.4$.

\section{Results}

Numerical results using the base case set of parameter values are presented below, followed by a number of sensitivity studies that provide additional insight into the cap formation process. We re-iterate here that we do not explicitly model the formation of the cap, but rather investigate the population dynamics of the collagen-synthesising SMCs. Our model is relevant during the early stages of fibrous cap formation, before the endothelium becomes overly distended by plaque growth and protrudes into the lumen. By comparing SMC volume fraction distributions in the intima under different parameter regimes, we use the model to assess the likely implications for fibrous cap stability.

\subsection{Base Case Simulation}

The initial SMC response to the PDGF profile in the base case simulation is shown in Figure~4: cells emerge from the media, migrate across the intima and begin to accumulate at the luminal boundary. SMCs continue to enter the intima over time, and an identifiable cap region, which displays a significantly elevated cell volume fraction, appears near the lumen after around 4--6 months (Figure~5a). This timescale shows excellent correlation with the work of \citet{Koza02}, who observed the development of fibrous caps in mice over periods of around 25 weeks. The increase in plaque SMC content, particularly near the lumen, reduces PDGF diffusion and increases uptake of PDGF. The PDGF gradient at the medial boundary also falls (Figure~5b), and the rate of recruitment of new cells is gradually reduced. SMCs continue to accumulate slowly, however, and the cell profile finally reaches a steady-state after around 15 months (Figure~5a). The total SMC content in the simulated plaque at this time is approximately 11.2\%, which is consistent with experimental data from studies using the atherosclerosis-prone apolipoprotein E (ApoE) knockout mouse \citep{Mall01, Clar06}.

	\begin{figure}[h!]
		\centering		
  		\includegraphics[width=0.48\textwidth]{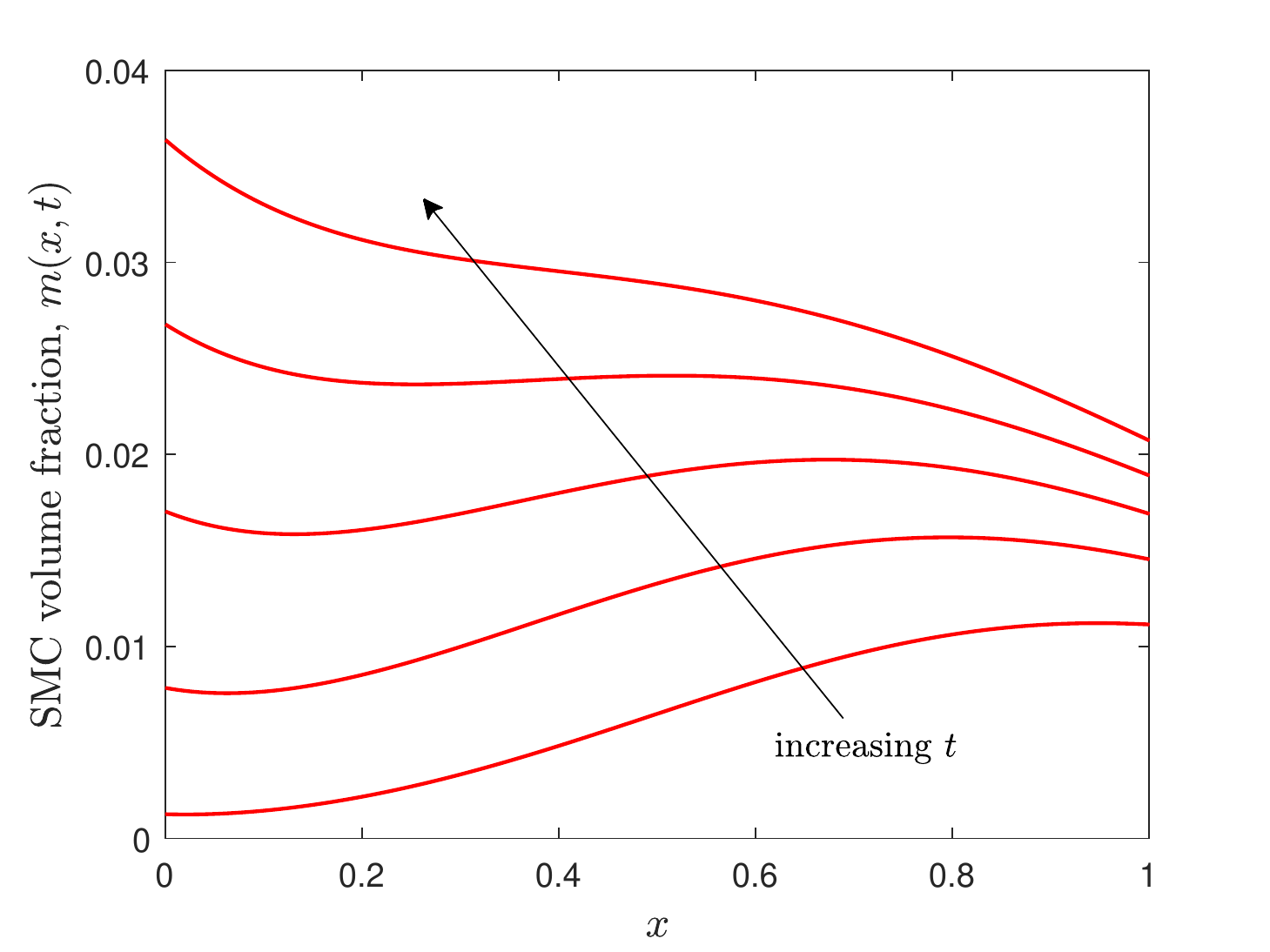}
		\caption{SMC volume fraction distributions in the intima at an early stage of cap development in the base case simulation.  The initial SMC volume fraction in the intima is small and uniform ($m\left(x,0\right)=1\times{10}^{-4}$).  SMCs enter the domain in response to the PDGF concentration gradient at the medial boundary ($x=1$) and migrate towards the PDGF source at the endothelium ($x=0$).  The arrow indicates the direction of sequential time points, which correspond to $t=$ 0.2, 0.4, 0.6, 0.8 and 1.}
	\end{figure}

	\begin{figure}[h!]
		\centering
		\begin{subfigure}[b]{0.48\textwidth}
			\includegraphics[width=\textwidth]{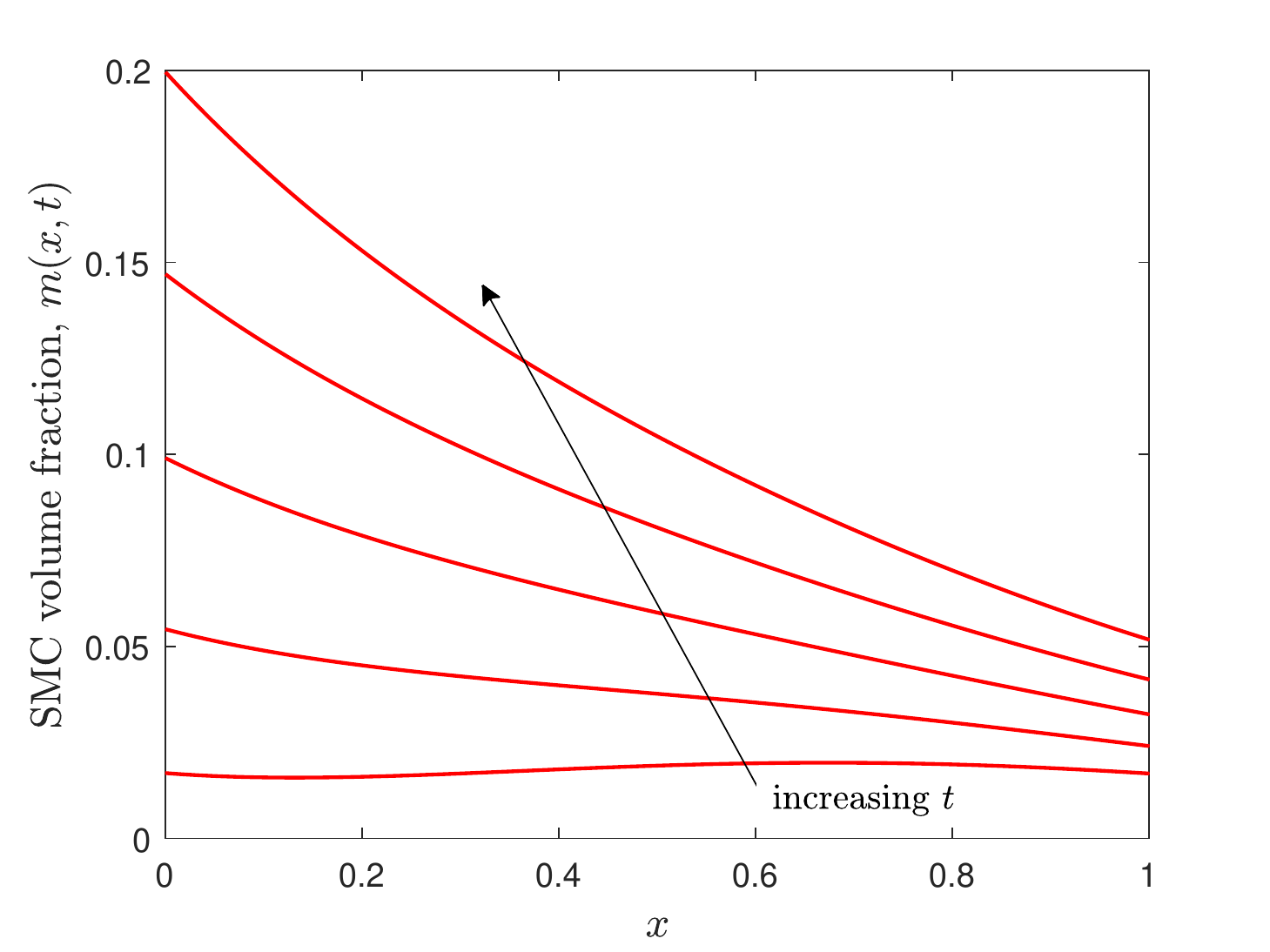}
			\caption{}
		\end{subfigure}
		\begin{subfigure}[b]{0.48\textwidth}
			\includegraphics[width=\textwidth]{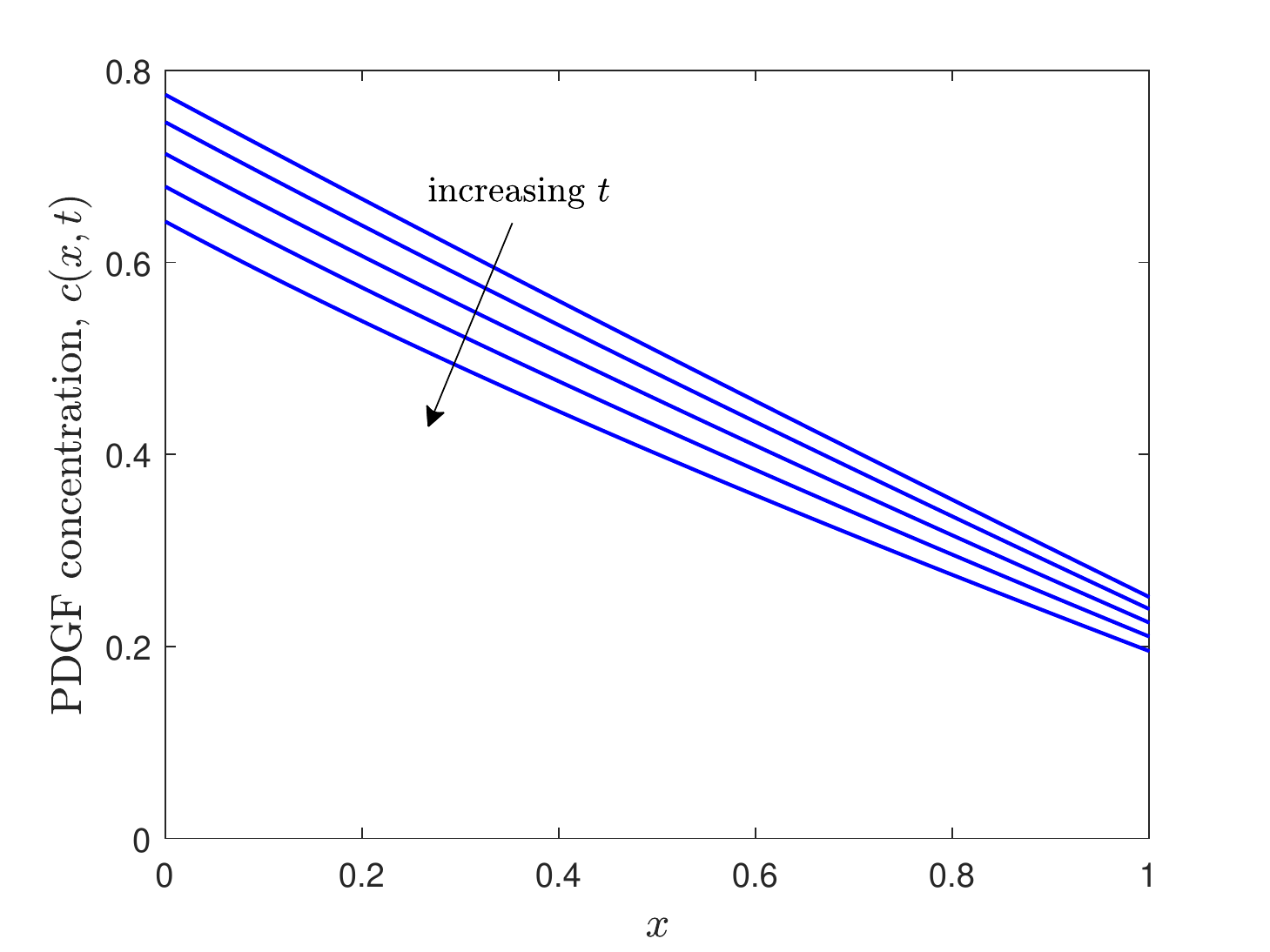}
			\caption{}
		\end{subfigure}
		\caption{(a) SMC volume fraction and (b) PDGF concentration distributions in the intima throughout cap development in the base case simulation.  SMC accumulation in the intima depletes the intimal PDGF due to increased PDGF uptake and reduced PDGF influx at the endothelium.  The arrows indicate the direction of sequential time points, which correspond to $t=$ 0.6, 1.4, 2.6, 4.6 and 15.  At the final time point, both variables have reached a steady-state.}
	\end{figure}

The above simulation demonstrates a dynamic interplay between the media-derived SMCs and the endothelium-derived PDGF that leads the SMC population to evolve naturally towards a profile that resembles a fibrous cap in the intima. Cap formation in the model arises from a highly nonlinear balance between SMC recruitment from the media, chemotactic migration towards the lumen and cell turnover through proliferation and apoptosis. However, the relative contribution of each of these factors, and their importance in determining overall cap stability, cannot be assessed from any single simulation. We therefore investigate each of these mechanisms in greater detail below by performing sensitivity simulations on a number of key model parameters.

\subsection{Sensitivity Analysis}

\subsubsection*{SMC Kinetics}

The formation of the fibrous cap requires the accumulation of matrix-synthesising SMCs near the endothelial lining of the arterial wall. Invasion from the media and proliferation within the intima are the two primary sources of these cells, but the extent to which each of these factors contributes to cap formation is unknown. Figure~6 compares the steady-state SMC profile from the base case with that from a simulation with no cell proliferation (i.e.\ $r=0$). As expected, proliferation contributes most strongly near the lumen where PDGF-stimulated cell division is largest. Overall, however, proliferation provides less than one quarter of the total SMC volume fraction in the base case simulation, and three quarters of the SMCs have migrated from the media. These results yield two interesting conclusions about fibrous cap synthesis. First, while SMC proliferation in the intima may enhance the thickness of a nascent cap, it does not appear to be necessary for cap formation. Second, the relative balance between SMC loss and SMC recruitment from the media (through the parameters $\beta_m$ and $m^*$, respectively) is likely to be a key factor in the ability of SMCs to form a stable cap. Sensitivity simulations on $\beta_m$ and $m^*$ support this latter conclusion, and also highlight that reducing SMC loss (i.e.\ reducing $\beta_m$) is a more effective way to increase SMC numbers in the cap region compared to increasing SMC recruitment (i.e.\ increasing $m^*$). This is because a reduction in SMC loss leads to a greater contribution from SMC proliferation, particularly near the endothelium where the PDGF signal is strong.

	\begin{figure}[h!]
		\centering		
  		\includegraphics[width=0.48\textwidth]{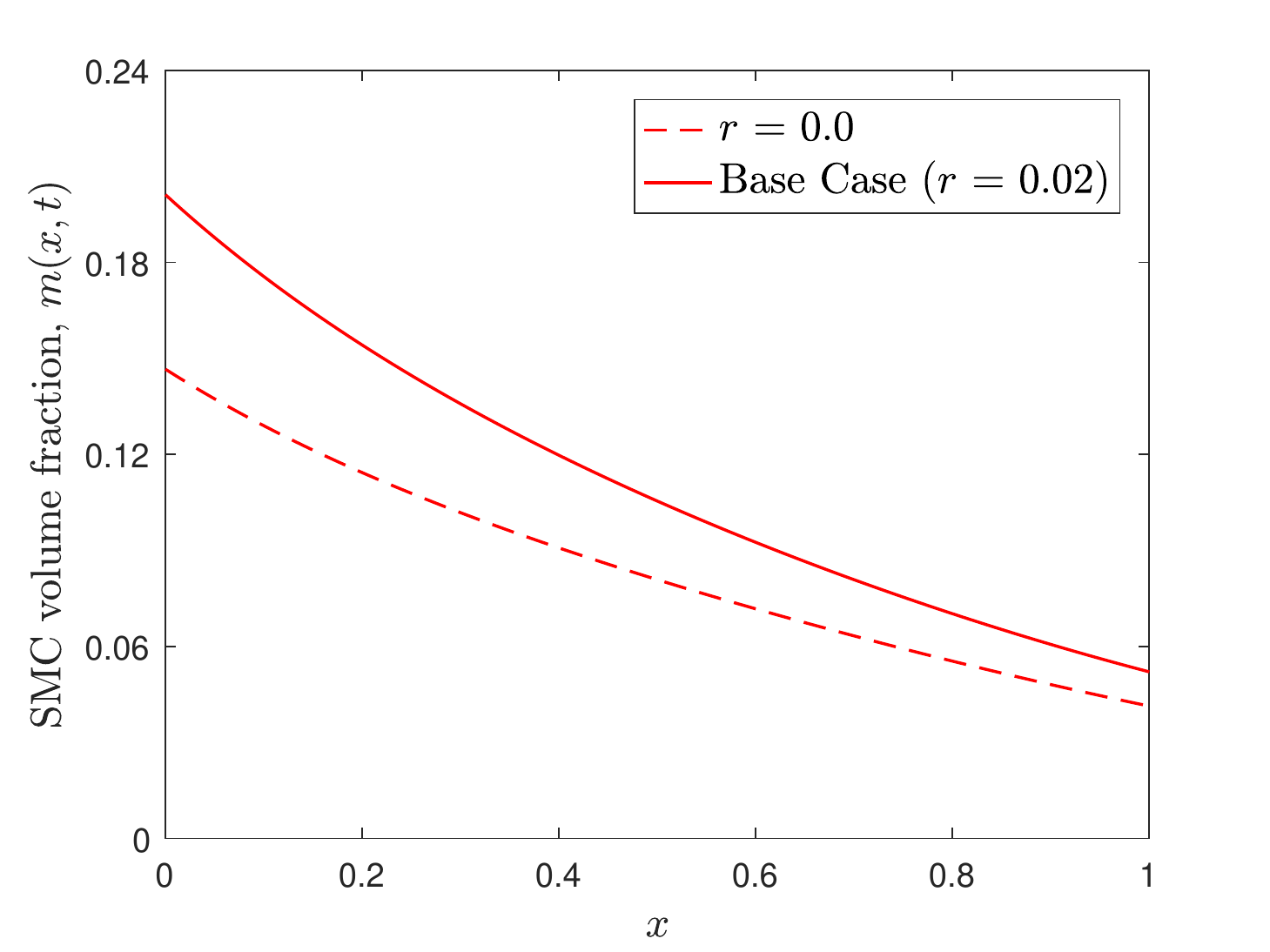}
		\caption{Steady-state SMC volume fraction distributions in the intima for the base case simulation (solid line) and a simulation with no SMC proliferation ($r=0$; dashed line).  Both plots are taken at time $t=24$.}
	\end{figure}

\subsubsection*{SMC Motility}

In our model, the recruitment of SMCs from the media and their subsequent migration across the intima is driven entirely by the PDGF profile. Therefore, it is important to study the parameters that control this migratory response and to determine the effect of SMC migration on cap formation. We investigate the impact of changing the exponent $n$ in the SMC phase pressure function $\Lambda\left(c\right)$, since this function provides an explicit link between SMC movement and the evolving chemical profile. We vary this parameter both upwards and downwards, choosing $n=1$ and $n=2.6$. The effect of these variations for both $\Lambda$ (i.e.\ diffusion) and $-\frac{d \Lambda}{d c}$ (i.e.\ chemotaxis) are plotted in Figure~7, and we emphasise here that the relevant range for dimensionless PDGF concentration during a typical simulation is around 0.1--1 (i.e.\ \numrange[range-phrase = --]{1}{10}~\si{\nano\gram\per\milli\litre}). In this range, the plots suggest that the general balance of the respective SMC flux terms should be tipped towards diffusion for $n=1$ and towards chemotaxis for $n=2.6$. This finding is confirmed by the steady-state SMC distributions for these scenarios (Figure~8).  For $n=1$, the model predicts a vast reduction in cell recruitment that leads to a shallow SMC profile with a barely discernible cap region. Alternatively, for $n=2.6$, the model yields a highly nonlinear SMC profile that potentially provides a stronger and more stable cap than the base case simulation. Interestingly, the overall SMC response is also much more efficient for $n=2.6$, reaching the improved steady-state profile in less than 12 months.

	\begin{figure}[h!]
		\centering		
  		\includegraphics[height=5cm]{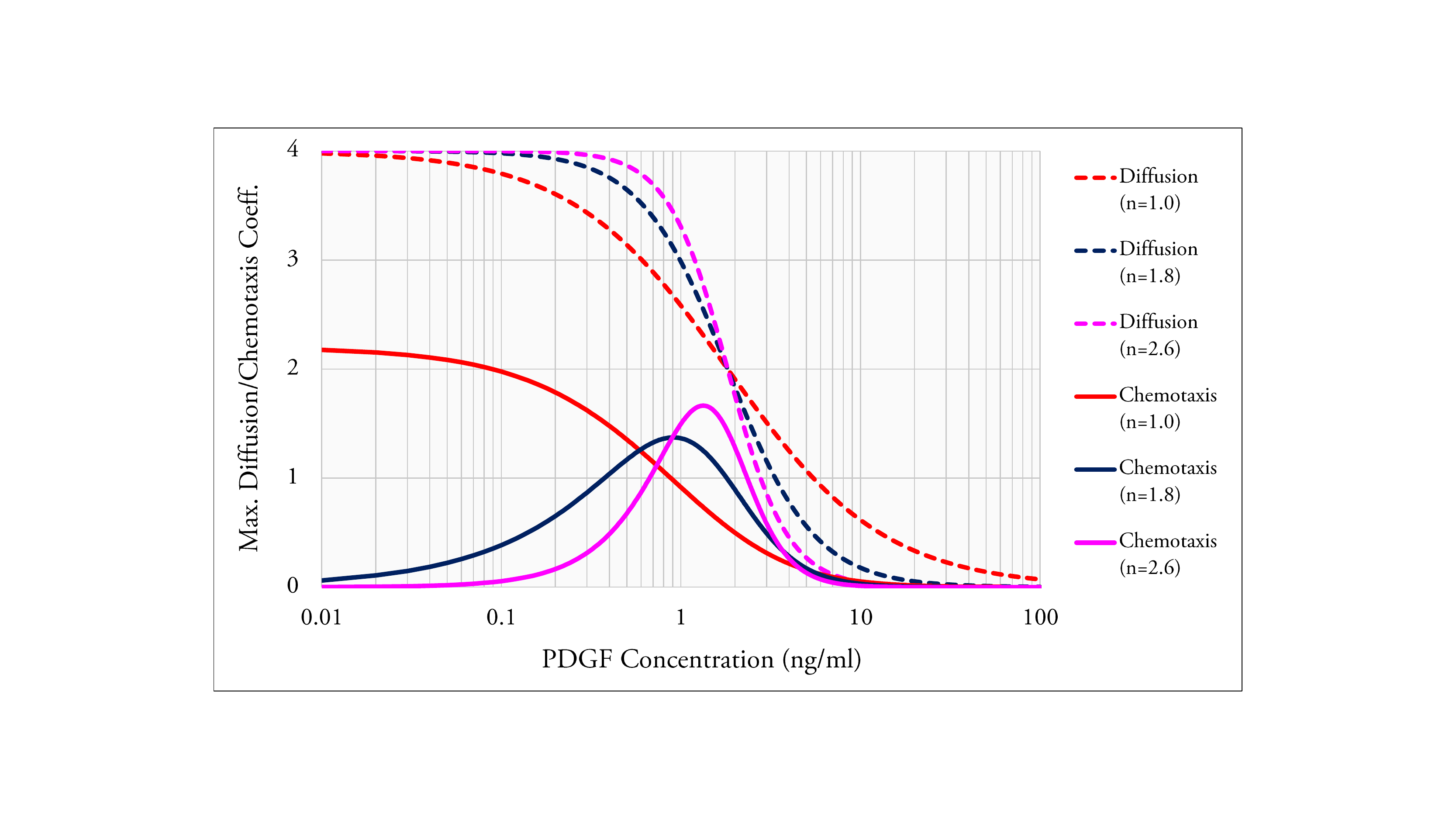}
		\caption{Sensitivity of the PDGF-dependent SMC diffusion (dashed lines) and chemotaxis (solid lines) coefficients to changes in the parameter $n$ in the SMC phase pressure function $\Lambda\left(c\right)$ (see equation (10)). Diffusion coefficients correspond to $\Lambda\left(c\right)$ and chemotaxis coefficients correspond to $-\frac{d \Lambda}{d c}$. The base case coefficients are shown in dark blue, coefficients for a reduced value ($n=1$) are shown in red, and coefficients for an increased value ($n=2.6$) are shown in magenta.}
	\end{figure}

	\begin{figure}[h!]
		\centering		
  		\includegraphics[width=0.48\textwidth]{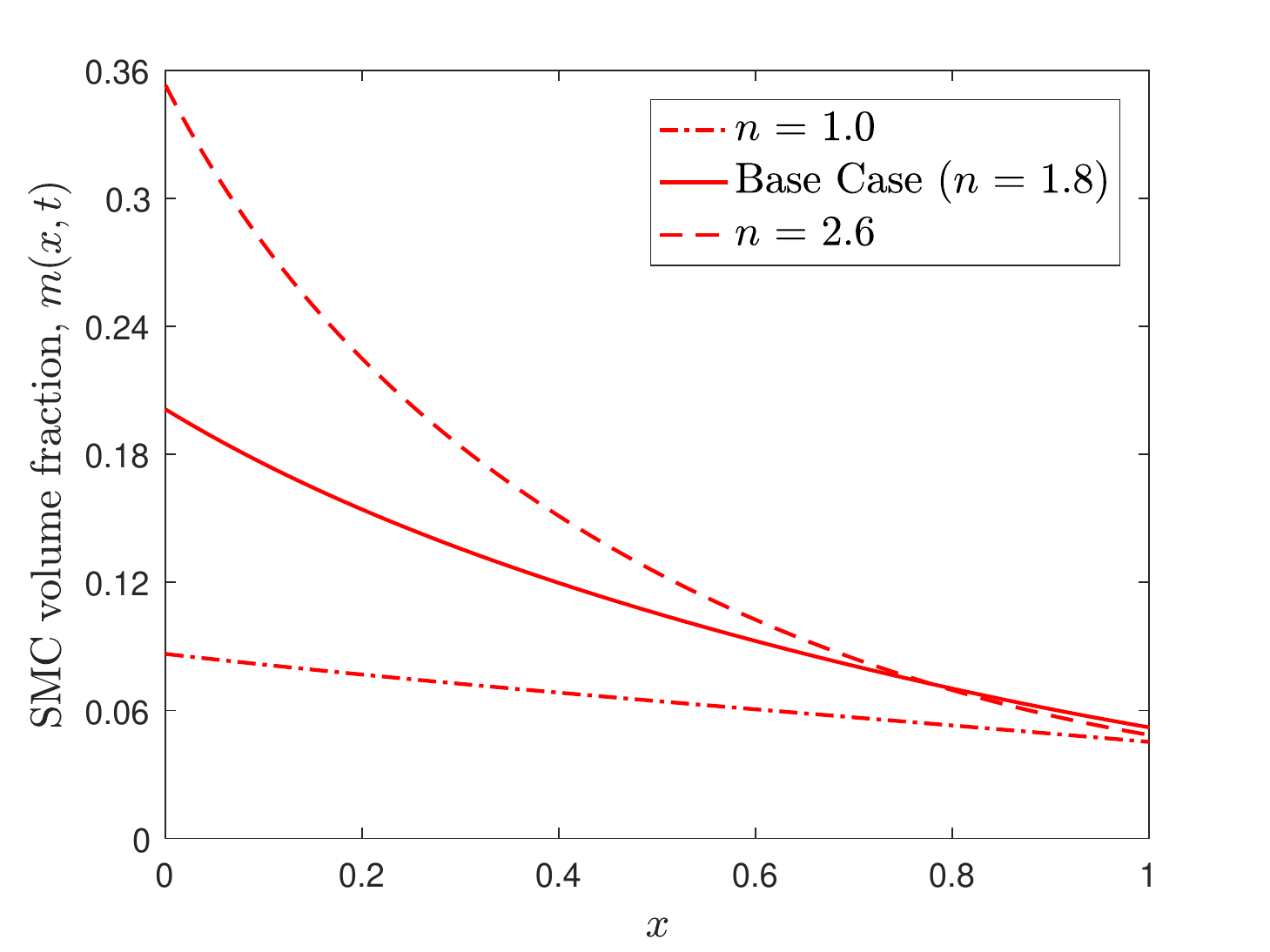}
		\caption{Simulation results showing how the steady-state SMC volume fraction distributions in the intima depend on the parameter $n$ in the SMC phase pressure function $\Lambda\left(c\right)$ (see equation (10)). Results from the base case simulation (solid line) are compared to results for a decrease ($n=1$; dot-dash line) and an increase ($n=2.6$; dashed line) in the value of $n$.  All plots are taken at time $t=24$.}
	\end{figure}

For completeness we briefly discuss our observations from varying the value of $\kappa$ in $\Lambda\left(c\right)$, although we do not present any results. Simulations indicate that the level of SMC recruitment to the cap region has a biphasic dependence on $\kappa$, where the maximum level of SMC recruitment occurs around the base case value ($\kappa=5.5$). In contrast to the results presented for $n$, cap formation appears to be relatively insensitive to changes in the value of $\kappa$. Values of $\kappa$ in the range 3.5 to 7.5, for example, produce only minor changes to the final steady-state SMC profiles. As $\kappa$ diverges from this range, cap formation is increasingly hampered, and, in the case of large $\kappa$ values, we note a significant reduction in the rate of formation of the compromised caps. This appears to be due to a combination of reduced SMC invasion from the media and a reduced rate of migration across the intima.

\subsubsection*{PDGF Profile}

In addition to manipulating the SMC response to PDGF, it is also of interest to explore the parameters controlling the evolution of the chemical profile itself. We consider the consequences of varying $\alpha_c$, which represents the rate of PDGF flux into the intima. We choose the values $\alpha_c=0.35$ and $\alpha_c=0.75$, and compare the initial PDGF profiles for these two scenarios and the base case ($\alpha_c=0.55$) in Figure~9. For $\alpha_c=0.35$ both the concentration and the gradient of PDGF is reduced throughout the intima, while for $\alpha_c=0.75$ both the concentration and the gradient of PDGF is increased throughout the intima. Note that these parameter values have been chosen to ensure that the chemical concentrations and gradients in the narrow intima remain within reasonable bounds. The SMC responses that result from these PDGF profiles are presented in Figure 10: the final steady-state SMC distributions are given in Figure~10a, while the overall SMC volume fractions in the intima (i.e.\ $\int_{0}^{1} m\left(x,t\right) \mathrm{d} x$) are plotted against time in Figure~10b. As would be expected, decreasing the production of PDGF at the lumen reduces the recruitment of SMCs from the media and reduces the potential for cap formation. Given the relatively low chemical gradient across the intima, however, the observed cellular response in this case is quick and remarkably robust. It appears likely that this is due to the pronounced peak in the chemotactic coefficient that occurs around dimensionless PDGF concentrations of 0.1 (i.e.\ \SI{1}{\nano\gram\per\milli\litre}; c.f.\ Figure 3).  This result contrasts strongly with the case where the influx of PDGF is increased. Despite the relatively steep PDGF gradient and the enhanced potential for SMC proliferation, this simulation predicts the slowest rate of SMC recruitment. Indeed, for $\alpha_c=0.75$ the total cell content in the intima lags behind that in the case with $\alpha_c=0.35$ for the first 7--8 months of the simulation. Moreover, increasing the value of $\alpha_c$ ultimately results in a cap that is less stable than the one generated in the base case. These results suggest that stimulating PDGF production has the potential to inhibit the SMC chemotactic response and thereby reduce both the efficiency and the efficacy of fibrous cap formation.

	\begin{figure}[h!]
		\centering		
  		\includegraphics[width=0.48\textwidth]{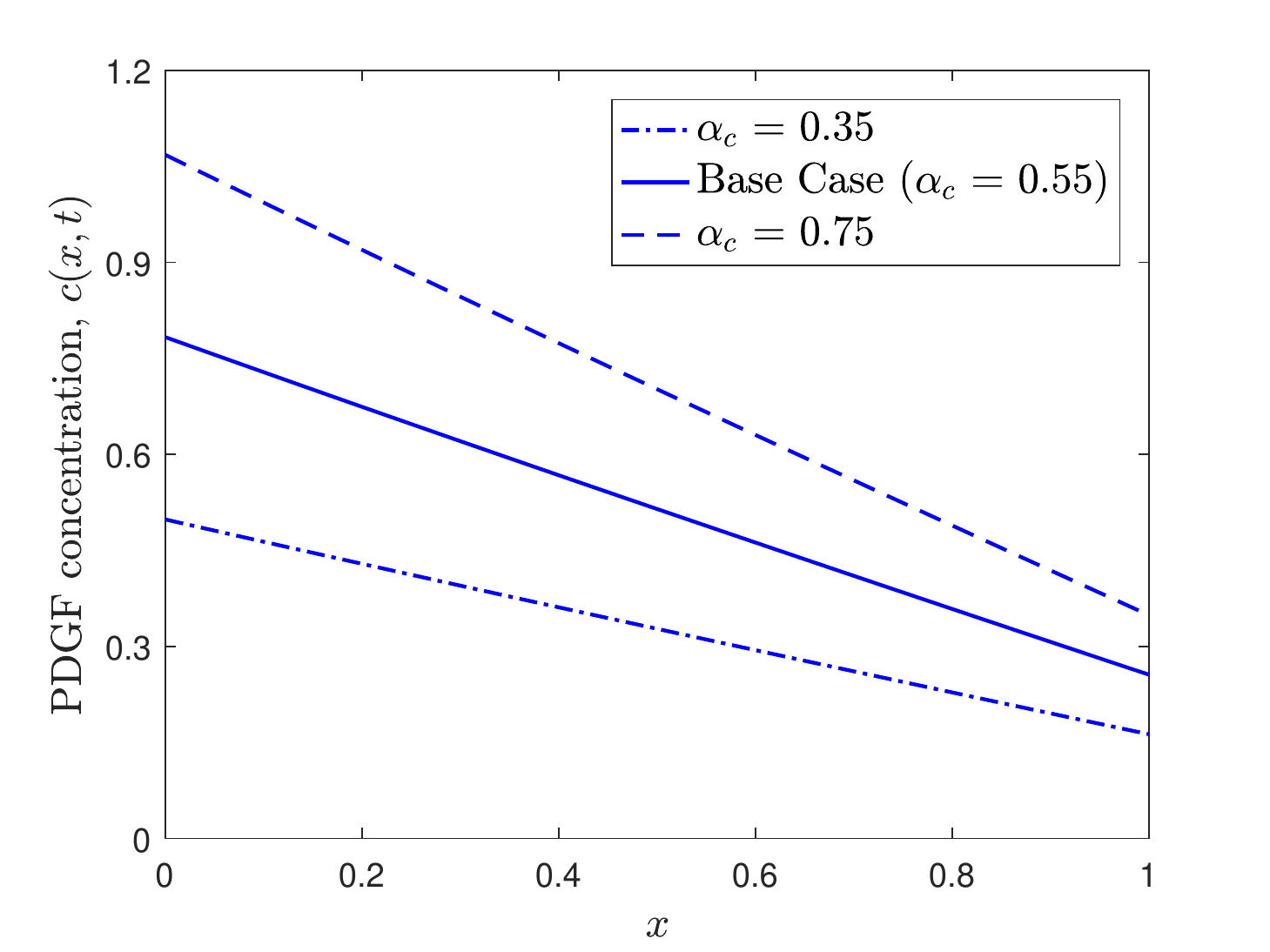}
		\caption{Simulation results showing how the initial PDGF concentration distribution in the intima changes as we vary the rate of PDGF influx from the endothelium $\alpha_c$. The initial PDGF profile for the base case simulation (solid lines) is compared to those for a decrease ($\alpha_c=0.35$; dot-dash line) and an increase ($\alpha_c=0.75$; dashed line) in the value of $\alpha_c$.}
	\end{figure}

	\begin{figure}[h!]
		\centering
		\begin{subfigure}[b]{0.48\textwidth}
			\includegraphics[width=\textwidth]{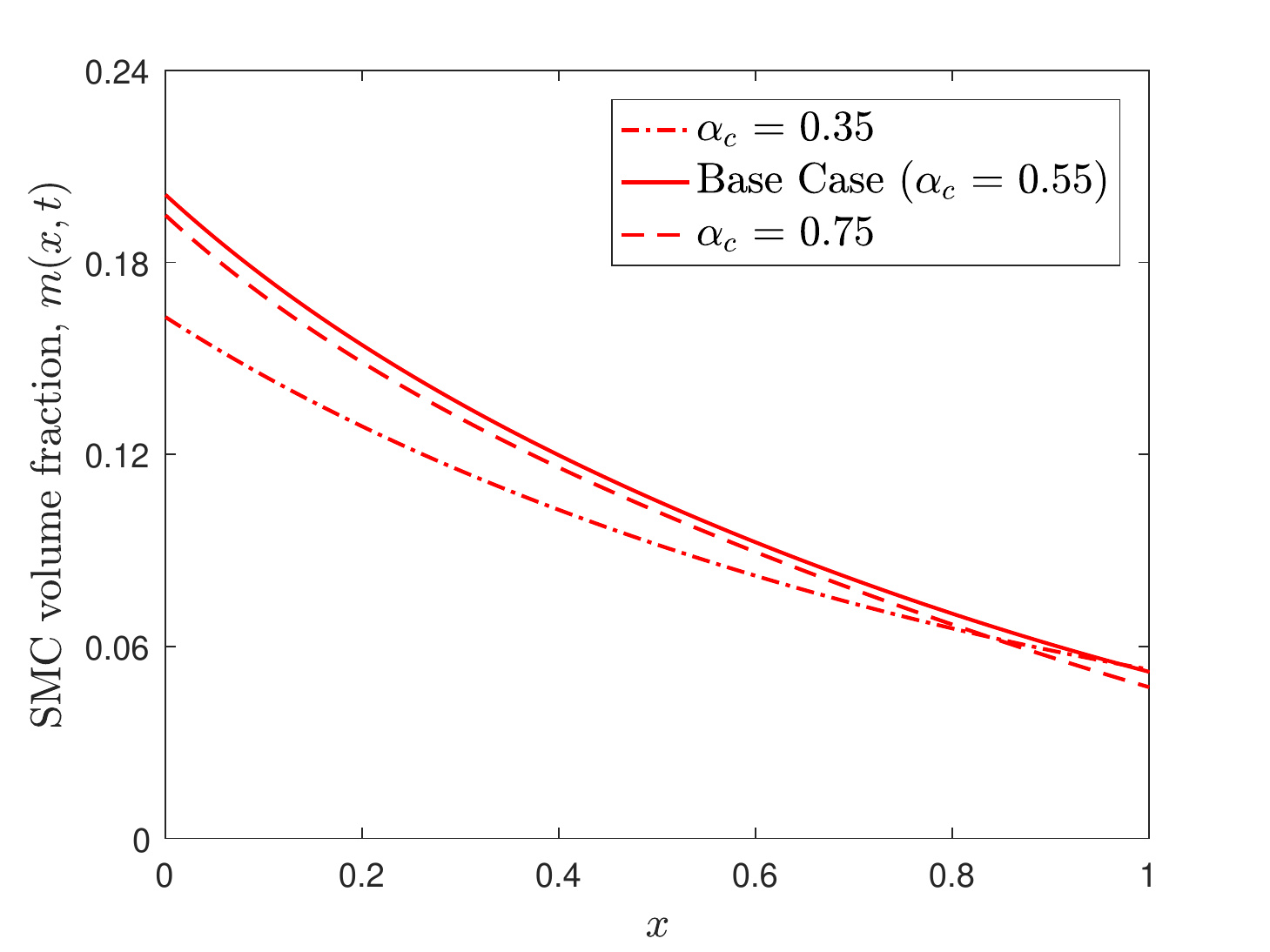}
			\caption{}
		\end{subfigure}
		\begin{subfigure}[b]{0.48\textwidth}
			\includegraphics[width=\textwidth]{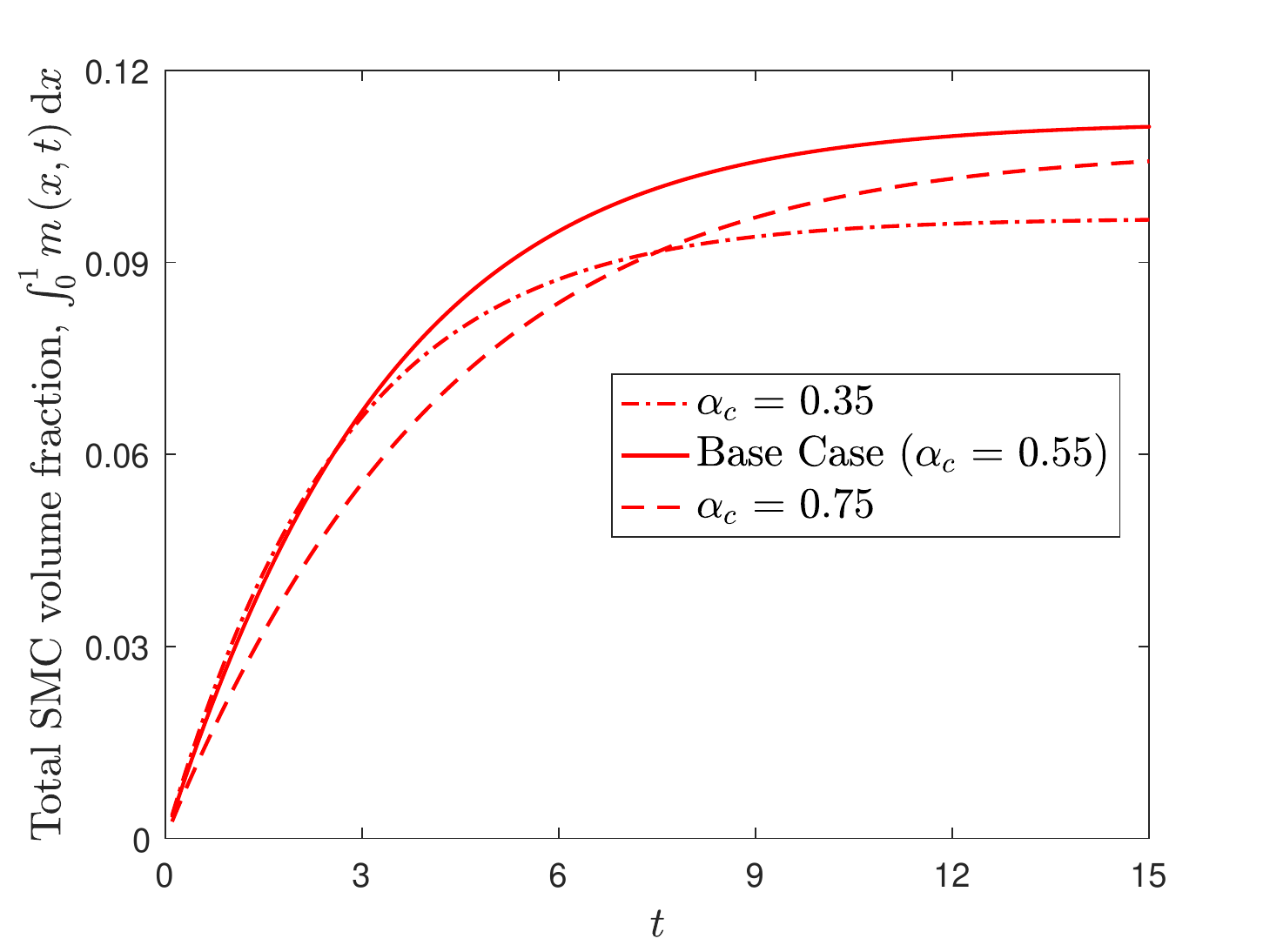}
			\caption{}
		\end{subfigure}
		\caption{Simulation results showing how the SMC response changes as we vary the rate of PDGF influx from the endothelium $\alpha_c$. The plots show (a) the final steady-state SMC volume fraction distribution in the intima and (b) the variation of the total SMC volume fraction in the intima over time for each simulation. Results from the base case simulation (solid lines) are compared to results for a decrease ($\alpha_c=0.35$; dot-dash line) and an increase ($\alpha_c=0.75$; dashed line) in the value of $\alpha_c$.  The steady-state plots are taken at time $t=24$.}
	\end{figure}

\section{Discussion}

In atherosclerosis, a protective fibrous cap is formed when SMCs are recruited from the media to synthesise new ECM and repair the damage caused by excessive accumulation of lipoproteins and inflammatory cells. The development of a stable fibrous cap is a key determinant of the prognosis for this disease, but the mechanisms of cap formation remain poorly understood. In this paper, we have formulated a multiphase model to investigate initial fibrous cap growth and identify key parameters that may contribute to cap stability. This work represents the first attempt to investigate fibrous cap formation in atherosclerosis using mathematical modelling. A key assumption of our approach has been to focus on SMCs only and not explicitly model collagen deposition. We have therefore modelled the dynamics of the collagen-synthesising SMC population, and assumed that the underlying matrix will evolve to broadly reflect the SMC profile in the diseased intima. This assumption has allowed us to derive a two-phase model that focusses on the chemotactic response of SMCs to endothelium-derived PDGF. We have used a variety of \emph{in vivo}, \emph{in vitro} and \emph{in silico} studies to obtain accurate estimates for key parameters in our model, and we have made assumptions that are consistent with observations of fibrous cap formation in the mouse. In our base case simulation, we have shown that both the timescale of cap formation and the total SMC volume fraction in the plaque are consistent with results from experiments using the atherosclerosis-prone ApoE knockout mouse.

The approach that we have taken in this paper has some limitations. For example, we have not explicitly considered the remodelling of the intimal tissue undertaken by SMCs recruited to \emph{in vivo} plaques. During cap formation, the initial elastin and proteoglycan matrix in the intima is degraded by SMCs and replaced by a new matrix primarily composed of fibrillar collagen. A range of \emph{in vivo} and \emph{in vitro} studies have suggested that haptotaxis in response to collagen --- as well as chemotaxis in response to growth factors such as PDGF --- may be an important mechanism of SMC migration in plaques \citep{Nels96, Lope13}. We have neglected this effect in the current study because we anticipate SMC chemotaxis will be the primary driver of fibrous cap formation, but in future studies we will investigate ECM remodelling in the plaque and the consequent impact of haptotaxis on SMC migration. Our fibrous cap model also does not include the inflammatory dynamics that take place in the plaque. Chronic plaque inflammation involves the sustained infiltration of lipoproteins and monocytes from the bloodstream, and the accumulation of fats and cellular debris in the intima. This process causes expansion of the intima during plaque development, and the implication is that fibrous cap growth may take place within tissue undergoing gradual expansion. We believe, however, that our fixed domain simulations of cap growth provide a reasonable initial approximation. More detailed approaches will be required to develop realistic coupled models of plaque inflammation and fibrous cap formation during atherosclerotic plaque growth.

Development of the fibrous cap (and the atherosclerotic plaque, in general) provides an intrinsically difficult challenge for mathematical modelling. The cap (or plaque) grows in the arterial intima, but much of this growth is driven by events that occur beyond this tissue layer. In the current context, for example, we have studied the activation and migration of medial SMCs in response to a PDGF signal generated at the endothelium. Our modelling domain represents only the narrow intima inside the artery wall, so the phenomena that occur at the endothelium and in the media have been captured by a set of non-standard boundary conditions.

For SMCs, the boundary condition imposed at the medial end of the domain (i.e.\ $x=1$) is crucial. Although the model does include a SMC proliferation term, most of the intimal SMCs are derived from the media and the model predicts no cap formation in the absence of this source. We have assumed that synthetic medial SMCs migrate through the IEL into the intima in an active (i.e.\ chemotactic) rather than a passive (i.e.\ diffusive) fashion. This assumption leads us to introduce the parameter $m^*$, which notionally represents a constant synthetic medial SMC volume fraction. In effect, however, $m^*$ implicitly quantifies a number of factors that we do not explicitly model, such as the rate of contractile to synthetic SMC transition and the rate of SMC migration in the medial tissue. Note that our modelling predicts the value of $m^*$ to be particularly low (i.e.\ $m^*=0.01$ for the base case simulation). This suggests that SMC recruitment from the media is likely to be a rate-limiting factor in fibrous cap development, which may reflect a physiological need to limit the depletion of SMCs in the media. The proliferation rate of SMCs in the media is believed to remain at low, non-diseased rates throughout atherosclerosis \citep{Lutg99}, so widespread SMC depletion in the media could ultimately have an adverse impact on cap stability.

In our model, the recruitment of SMCs from the media to form the fibrous cap is reliant upon a gradient of PDGF. In the absence of a PDGF gradient at the medial boundary, no SMCs will enter the domain from the media and a cap of SMCs at the endothelium will not be formed. We assume there is a flux of PDGF into the intima from the damaged endothelium (i.e.\ at $x=0$), and that a gradient of PDGF at the medial boundary is created by outward diffusion of PDGF into the media through the porous IEL. The width of the domain on which we solve our equations (\SI{75}{\micro\metre}) is relatively small compared to the PDGF diffusion distance (\SI{300}{\micro\metre}). The model therefore typically predicts PDGF profiles that are approximately linear with significant loss of PDGF to the media. This is reasonable provided we assume, for example, that SMCs in the media uptake significant PDGF. In general, however, it is likely that our intimal PDGF profiles may not be as well-developed as they would be on a larger domain. This factor was taken into account when we made our parameter selections, and we have only considered PDGF profiles that demonstrate biologically reasonable gradients and concentrations. We will address this limitation in future studies by developing models on larger domains that include both the intima and the media, but we do not anticipate any significant changes to the broad conclusions of the current work.

The intimal SMC profiles generated by our model are the result of a dynamic combination of SMC diffusion, chemotaxis and proliferation --- all of which depend nonlinearly on the PDGF concentration. For both chemotaxis and proliferation, the model responses have been informed by \emph{in vitro} measurements on vascular SMCs. SMC chemotaxis shows a biphasic response to PDGF across a range of intermediate PDGF concentrations. SMC proliferation, on the other hand, tends to occur at larger PDGF concentrations where the chemotactic sensitivity is significantly diminished. This pattern of SMC behaviour is consistent with current understanding of fibrous cap formation: PDGF produced at the damaged endothelium recruits SMCs by chemotaxis from the distal media and then stimulates their proliferation to enhance matrix synthesis and tissue stability proximal to the cap region. Our simulations are consistent with this theory, but they have also identified several parameters that can influence the effectiveness of the cap formation process.

A key model prediction is that the influence of SMC proliferation appears to be relatively small during cap formation. At steady-state in the base case simulation, for example, we identified that only 25\% of the SMCs in the cap region could be attributed to mitosis, while the remaining 75\% had been recruited from the media. Although the rates of SMC proliferation in our simulations are rather low (doubling times range from about 3 months to 18 months depending on the local PDGF concentration), the limited contribution from proliferating cells also appears to be a consequence of the narrow width of the intima. Across this thin tissue region, we can only expect relatively shallow PDGF gradients. As a result, the PDGF concentration in the cap region is generally similar to that at the medial boundary. Where these PDGF concentrations are low enough to favour chemotactic SMC recruitment from the media, proliferation rates in the cap region are suppressed. Alternatively, where the PDGF concentrations are high enough to favour proliferation, SMC recruitment is reduced and so is the pool of cells available for division. Of course, these results are based on the assumption of fixed PDGF influx from the endothelium. If PDGF production was gradually increased over time, for example, we may observe significant SMC migration to the cap region at low PDGF followed by significant SMC proliferation at high PDGF, ultimately leading to enhanced cap stability. These ideas raise interesting questions as to how the balance between SMC migration and proliferation may evolve as the intima expands during long-term plaque development, and also how this balance may translate to human atherosclerosis where the intima is typically much wider than in the mouse.

As a consequence of the limited potential for SMC mitosis in our model, the dynamic balance between between SMC recruitment from the media (controlled by $m^*$) and SMC loss within the plaque (controlled by $\beta_m$) becomes critical to the formation of a stable fibrous cap. The rate of SMC migration (controlled by the cell motility coefficient $\chi$) is also important to cap formation, but note that the impact of an increase in the migration rate would be limited if medial SMCs were in short supply (i.e.\ for very small $m^*$). We have not presented the results here, but our simulations indicate that the total SMC content in the intima is strongly sensitive to variations in the values of both $\beta_m$ and $m^*$. In the particular case of an increase in $\beta_m$, we observe reduced SMC numbers in the cap region and anticipate that this would adversely affect cap stability. This interpretation is consistent with the experimental results of \citet{Clar06}. By developing a transgenic mouse in which plaque SMCs could be induced to undergo apoptosis by treatment with diphtheria toxin, \citet{Clar06} found that reduced plaque collagen content and reduced fibrous cap thickness were direct consequences of elevated SMC death.

The dynamic balance between SMC diffusion and chemotaxis may also play a crucial role in the fibrous cap formation process. Within the framework of our model, these two forms of cell movement are highly nonlinear and intrinsically linked by the SMC phase pressure function $\Lambda\left(c\right)$. The SMC diffusion coefficient is directly proportional to $\Lambda$ itself, while the SMC chemotaxis coefficient is instead proportional to $-\frac{d \Lambda}{d c}$. The function $\Lambda\left(c\right)$ is characterised by the parameters $n$ and $\kappa$, and both of these influence the SMC response to PDGF signalling. In terms of SMC recruitment to the cap region, our simulations were insensitive to $\kappa$ across a range of values from 3.5 to 7.5. The exponent $n$, on the other hand, had a dramatic impact on SMC cap formation. Within the relevant range of PDGF concentration, decreasing $n$ decreases the chemotaxis coefficient and increases the diffusion coefficient, while increasing $n$ has the opposite effect. Consequently, a simulation with reduced $n$ produced a shallow SMC profile that lacked a discernible cap region, while a simulation with increased $n$ produced a steep SMC profile with an enhanced SMC volume fraction in the cap region. These features of the SMC profile in the large $n$ simulation are highly desirable in the context of plaque stability. The results in this case display an abundance of SMCs near the endothelium and relatively few beyond, which implies that the plaque will form a dense, protective cap region without excessive hardening of the remainder of the artery. These results highlight the importance of a strong SMC chemotactic response during fibrous cap formation.

The evolution of PDGF concentration in our simulations is influenced by factors that include the uptake of PDGF by migrating SMCs and modulated diffusive transport of PDGF due to SMC accumulation near the endothelium. However, the PDGF profile in the intima is primarily determined by the influx and efflux of PDGF, which depend on the parameters $\alpha_c$ and $\sigma_c$, respectively. Sensitivity simulations for $\alpha_c$ produce surprising results. Steady-state SMC profiles were found to be relatively insensitive to $\alpha_c$ values between 0.35 and 0.75, despite a greater than two-fold increase in the intimal PDGF gradient over this range. Moreover, the \emph{steepest} PDGF gradient yielded the \emph{slowest} rate of cap formation. This result reflects the fact that an increase in the chemical gradient was associated with a concurrent increase in the PDGF concentration levels throughout the domain. The elevated PDGF levels significantly reduced the chemotactic sensitivity of SMCs to PDGF, and this slowed cap formation by reducing both the rate of SMC recruitment from the media and the rate of SMC migration in the intima. The fact that our model can predict the formation of stable fibrous caps over a range of timescales is significant because it reflects the wide variation in the observed times required for thick cap formation in experimental studies \citep{Koza02}. These results indicate that, while a gradient of PDGF in the intima is necessary for cap formation, the stability of the fibrous cap and the rate of cap formation may depend critically upon the maintenance of appropriate PDGF concentration levels.

In the biomedical literature there is a prevailing view that a growing atherosclerotic plaque will establish an initially strong fibrous cap, but plaque rupture may ultimately be induced by a gradual thinning or erosion of the cap over time. It is important to emphasise that the model presented in this paper focusses purely on the early process of cap establishment, and does not include factors that may become relevant for long-term cap maintenance. Mechanisms of fibrous cap erosion are poorly understood, but a number of factors --- many of which can be linked to plaque inflammation --- have been proposed. These include excessive matrix degradation by macrophage-produced MMPs \citep{Lusi00} and reduced SMC matrix production due to a phenotypic change that has been linked to excessive lipoprotein consumption by SMCs in plaques \citep{Benn16}. Plaque rupture due to fibrous cap thinning is known to be a pre-cursor to heart attacks and strokes. Modelling mechanisms of cap erosion will be a key target for future studies.

In this work we have developed a multiphase model to investigate the response of medial SMCs to endothelium-derived PDGF signalling in atheroscl-\\erotic plaques. Our model predicts that SMCs will accumulate near the endothelium and form fibrous caps for a wide range of parameter sets, but also that variations in key parameters can lead to changes in cap thickness. In particular, we have found cap thickness to be particularly sensitive to the balance between SMC apoptosis and SMC recruitment from the media, as well as to the balance between SMC diffusion and SMC chemotaxis. We have also demonstrated that the rate of cap formation can be highly variable, and that stable caps can form over a wide range of time scales. This study represents the foundation of a mechanistic modelling framework that can be readily extended to further investigate mechanisms of cap formation and atherosclerotic plaque development in general.

\section*{Acknowledgements}
MGW, CM and MRM acknowledge funding from an Australian Research Council Discovery Grant (DP160104685).



\end{document}